%Paper: hep-ph/9501361
%From: Shyamoli Chaudhuri <sc@itp.ucsb.edu>
%Date: Mon, 23 Jan 1995 17:09:04 -0800
%Date (revised): Thu, 31 Aug 95 11:43:50 PDT

\input harvmac
\Title{\vbox{\baselineskip12pt\hbox{FERMILAB-PUB-94/413-T}
\hbox{NSF-ITP-94-71}}}
{\vbox{\centerline{String Consistency for Unified Model Building}}}

\centerline{{\bf S. Chaudhuri\footnote{$^\dagger$}
{e-mail : sc@itp.ucsb.edu},
S.-w. Chung$^*$, G. Hockney$^*$, and J. Lykken\footnote{$^*$}
{e-mail : chung@fnth04.fnal.gov, hockney@fnalv.fnal.gov, and
lykken@fnalv.fnal.gov}}}
\medskip\centerline{$^{\dagger}$Institute for Theoretical Physics,}
\centerline{University of California, Santa Barbara, CA 93106}
\medskip\centerline{$^{*}$Theory Dept., MS106,}
\centerline{Fermi National Accelerator Laboratory,}
\centerline{P.O. Box 500, Batavia, IL 60510}
\vskip .8in
\centerline{\bf ABSTRACT}
We explore the use of real fermionization as a test case
for understanding how specific features of phenomenological
interest in the low-energy effective superpotential
are realized in exact solutions to heterotic superstring theory.
We present pedagogic examples of models which realize SO(10)
as a level two current algebra on the world-sheet, and discuss
in general how higher level current algebras can be realized
in the tensor product of simple constituent conformal field
theories. We describe formal developments necessary to compute
couplings in models built using
real fermionization.
This allows us to isolate cases of
spin structures where the standard
prescription for real fermionization may break down.

\Date{12/94}
%%%%%%%%%%%%%%%%
%% local macros
%%
\catcode`\@=11 % This allows us to modify PLAIN macros.
\def\mymatrix#1{\null\,\vcenter{\normalbaselines\m@th
  \ialign{\hfil$##$&&\quad\hfil$##$\crcr
	\mathstrut\crcr\noalign{\kern-\baselineskip}
	#1\crcr\mathstrut\crcr\noalign{\kern-\baselineskip}}}\,}
\catcode`\@=12 % at signs are no longer letters
\global\newcount\fno \global\fno=0
\def\myfoot#1{\global\advance\fno by1
\footnote{$^{\the\fno}$}{#1}}
\def\chit{{\tilde\chi}}
\def\ru{\hrulefill}
\def\vr{\vrule}

\def\br{\bar{N}_{s}}
\def\bi{\bar{N}_{c}}

\def\ket#1{\left| #1\right\rangle}
\def\i{{\cal I}}
\def\vev#1{\langle #1 \rangle}
\def\N{N_{ijk}}

\def\c{{\rm cos}}
\def\s{{\rm sin}}

\def\k{k_{ij}}

\def\cmp#1{{\it Comm. Math. Phys.} {\bf #1}}
\def\pl#1{{\it Phys. Lett.} {\bf #1B}}
\def\prl#1{{\it Phys. Rev. Lett.} {\bf #1}}
\def\prd#1{{\it Phys. Rev.} {\bf D#1}}

\def\np#1{{\it Nucl. Phys.} {\bf B#1}}

\def\p{{\scriptscriptstyle +}}
\def\m{{\scriptscriptstyle -}}
\def\be{\begin{equation}}
\def\ee{\end{equation}}

\def\zbar{\bar z}

\def\darr#1{\raise1.5ex\hbox{$\leftrightarrow$}\mkern-16.5mu #1}
 %pound sterling
\def\half{{1\over2}}

\def\pmsq{\pm {\textstyle{1\over{\sqrt 2}}}}

\def\ha{{{\scriptstyle 1}\over{\scriptstyle 2}}} %puts a small half in math
 %puts a small sixteenth in a displayed eqn
\def\roughly#1{\raise.3ex\hbox{$#1$\kern-.75em\lower1ex\hbox{$\sim$}}}
\def\c{{\rm cos}}
\def\s{{\rm sin}}

\def\k{k_{ij}}
\def\e#1{{\rm e}^{#1}}
\def\p{\raise2pt\hbox to5pt{\fiverm +}}
\def\m{\raise2pt\hbox to5pt{\fiverm --}}
\def\fm{\m\m\m\m}

\def\cmp#1{{ Comm. Math. Phys.} {\bf #1}}
\def\pl#1{{ Phys. Lett.} {\bf #1B}}
\def\prl#1{{ Phys. Rev. Lett.} {\bf #1}}
\def\prd#1{{ Phys. Rev.} {\bf D#1}}

\def\np#1{{ Nucl. Phys.} {\bf B#1}}

\def\be{\begin{equation}}
\def\ee{\end{equation}}

\def\taubar{\bar \tau }

\def\zbar{\bar z}

\def\vev#1{\langle #1 \rangle}

\def\darr#1{\raise1.5ex\hbox{$\leftrightarrow$}\mkern-16.5mu #1}
 %pound sterling
\def\half{{1\over2}}

\def\pmsq{\pm {\textstyle{1\over{\sqrt 2}}}}

 %puts a small sixteenth in a displayed eqn
 %puts a small three-half in a displayed eqn
 %puts a small 1/48 in a displayed eqn
\def\roughly#1{\raise.3ex\hbox{$#1$\kern-.75em\lower1ex\hbox{$\sim$}}}
%%%%%%%%%%%%%%%%%%%%%%%%%%%%%%%%%%%%%%%%%%%%%%%%%%%%%%%%%%%%%%%%%%%%%%%%%%
%references
\lref\rametal{M. Leurer, Y. Nir, and N. Seiberg, \np{398} (1993) 319;
L. Ibanez and G. Ross, \pl{332} (1994) 100;
D. Kaplan and M. Schmaltz, \prd{49} (1994) 3741;
P. Binetruy and P. Ramond, {\it Yukawa textures and anomalies},
Inst. for Fundamental Theory preprint UFIFT-HEP-94-19.}
\lref\ilr{I. Antoniadis, J. Ellis, S. Kelley, and D. Nanopoulos,
\pl{272} (1991) 31;
L. Ibanez and D. Lust, \pl{272} (1991) 251;
M. Gaillard and R. Xiu, \pl{296} (1992) 71.}
\lref\textures{J. Ellis and M. Gaillard, \pl{88} (1979) 315; S. Dimopoulos,
\pl{129} (1983) 417; D. Nanopoulos and M. Srednicki,
\pl{124} (1983) 37;
J. Bagger, S. Dimopoulos, H. Georgi, and S. Raby, Proc. of the Fifth Workshop
on Grand Unification, Providence, Rhode Island (1984).}
\lref\fqs{D. Friedan, Z. Qiu, and S. Shenker, \prl{52} (1984) 1575.}
\lref\strom{E. Witten, \np{268} (1986) 79; A. Strominger, \np{274}
(1986) 253.}
\lref\gins{P. Ginsparg, \pl{197} (1987) 139.}
\lref\adhrs{G. Anderson, S. Dimopoulos, L. Hall, S. Raby, and G. Starkman,
\prd{49} (1994) 3660;
S. Raby, {\it SO(10) SUSY GUTs and fermion masses}
Ohio State preprint OHSTPY-HEP-T-94-010.}
\lref\far{A. Faraggi, \pl{274} (1992) 47; \pl{278} (1992) 131;
\pl{302} (1993) 202; \pl{326} (1994) 62; A. Faraggi, D. Nanopoulos
and K. Yuan, \np{335} (1990) 347.}
\lref\lnyb{J. Lopez, D. Nanopoulos and K. Yuan, \prd{50} (1994) 4060.}
\lref\lnya{J. Lopez, D. Nanopoulos and K. Yuan, \np{399} (1993) 654.}
\lref\dreiner{H. Dreiner, J. Lopez, D. Nanopoulos and D. Reiss,
\np{320} (1989) 401.}
\lref\cs{S. Chaudhuri and J. Schwartz, \pl{219} (1989) 291.}
\lref\babu{K. S. Babu and S. M. Barr, \prd{51} (1995) 2463; \prd{49}
(1994) 2156.}
\lref\dw{S. Dimopoulos and F. Wilczek, Proc. of Erice Summer School (1981).}
\lref\moha{K. S. Babu and R. N. Mohapatra, \prl{74} (1995) 2418.}
\lref\nelson{M. Dine, A. Nelson, and Y. Shirman, {\it Low energy
dynamical supersymmetry breaking
simplified}, Santa Cruz preprint SCIPP 94/21 (1994).}
\lref\dine{See, for example, M. Dine, P. Huet and N. Seiberg, \np{322}
(1989) 301; D. Bailin, D. Dunbar and A. Love, \pl{219} (1989) 76;
J. Casas and C. Munoz, \np{332} (1990) 189;
L. Ibanez and G. Ross, \np{368} (1992) 3.}
\lref\rad{L. Alvarez-Gaume, M. Claudson and M. Wise, \np{207} (1982) 16;
J. Ellis, L. Ibanez, and G. Ross, \pl{113} (1982) 283;
L. Alvarez-Gaume, J. Polchinski and M. Wise,
\np{221} (1983) 495; L. Ibanez, \np{218} (1983) 514; J. Ellis,
J. Hagelin, D. Nanopoulos,
and K. Tamvakis, \pl{125} (1983) 275.}
\lref\guts{E. Witten, \np{188} (1981) 513; S. Dimopoulos and H. Georgi,
\np{193} (1982) 475;
N. Sakai, Z. Phys. C11 (1982) 153. For an introduction, see G. Ross,
{\it Grand Unified Theories}, Benjamin, New York, 1984.}
\lref\ss{T. Yoneya, {\it Prog. Theo. Phys.} {\bf 51} (1974) 1907;
J. Scherk and J. H. Schwarz, \np{81} (1974) 118.}
\lref\gso{F. Gliozzi, J. Scherk, and D. Olive, \pl{65} (1976) 282;
\np{122} (1977) 253.}
\lref\rns{P. Ramond, \prd{3} (1971) 2415;  A. Neveu and J. H.
Schwarz, \np{31} (1971) 86;
K. Bardakci and M. Halpern, \prd{3} (1971) 2493.}
\lref\kln{S. Kalara, J. Lopez and D. Nanopoulos, \np{353} (1991) 650.}
\lref\gs{M. B. Green and J. H. Schwarz, \pl{136} (1984) 367; \pl{149}
(1984) 117.}
\lref\st{M. B. Green, J. H. Schwarz, and E. Witten, {\it Superstring Theory},
Cambridge, Cambridge (1987). For a recent review of conformal field theory
and string theory, see J. Polchinski, {\it What
is String Theory?}, Lectures in Les Houches LXII, hep-th/9411028 (1994).}
\lref\chsw{P. Candelas, G. Horowitz, A. Strominger and E. Witten, \np{258}
(1985) 46. For an introduction, see E. Witten, {\it Grand
Unification in Four or Ten Dimensions}, Sixth Workshop on Grand Unification,
ed. S. Rudaz and T. Walsh (1985).}
\lref\shap{J. A. Shapiro, \prd{5} (1972) 1945;
J. Polchinski, \cmp{104} (1986) 37.}
\lref\sw{N. Seiberg and E. Witten, \np{276} (1986) 272.}
\lref\us{S. Chaudhuri, S. Chung and J. Lykken, {\it
Fermion masses from superstring models with adjoint scalars},
hep-ph/9405374, May 1994;
S. Chaudhuri, S. Chung, G. Hockney and J. Lykken, Proc. of
the DPF meeting, Albuquerque, NM, hep-th/9409151, August 1994.}
\lref\kac{V. G. Kac, {\it Funct. Anal. App.}{\bf 1} (1967) 328; R. V. Moody,
{\it Bull. Am. Math. Soc.}{\bf 73} (1967) 217.}
\lref\dhl{ D. Lewellen, \np{337} (1990) 61.}
\lref\patera{W. McKay and J. Patera, {\it Tables of dimensions, indices
and branching rules for representations of simple algebras}
(Dekker, New York, 1981).}
\lref\fer{A. Font, L. Ibanez, and F. Quevedo, \np{345} (1990) 389.}
\lref\klt{H. Kawai, D. C. Lewellen, and S.-H. H. Tye,
\np{288} (1987) 1.}
\lref\docspec{The symbolic manipulation package
will be described and documented in a future publication.
Additional information is available on the World-Wide Web at
http://www-theory.fnal.gov/superstrings/superstrings.html}
\lref\ibanez{G. Aldazabal, A. Font, L. Ibanez, A. Uranga, {\it String
GUTS}, FTUAM-94-28, hep-th@xxx.lanl.gov-9410206; L. Ibanez,
in Susy'94, Ann Arbor (May 1994).}
\lref\abk{I. Antoniadis, C. P. Bachas, and C. Kounnas,
\np{289} (1987) 87; I. Antoniadis and C. P. Bachas, \np{298} (1988) 586.}
\lref\klst{H. Kawai, D. C. Lewellen, J. A. Schwartz, and S.-H. H. Tye,
\np{299} (1988) 431.}
\lref\joel{J. A. Schwartz, \prd{42} (1990) 1777.}
\lref\ffont{A. Font, L. Ibanez, F. Quevedo, and A. Sierra,
\np{331} (1990) 421.}
\lref\dix{For an introduction, see L. Dixon, {\it Some world-sheet
properties of
superstring compactifications}, in the ICTP Summer Workshop, Trieste (1987)
QCD161:W626:1987; {\it ibid.}, {\it Supersymmetry breaking in string theory},
in the DPF meeting, Houston TX (1990) QCD161:T45:1989.}
\lref\dkl{L. Dixon, V. Kaplunovsky and J. Louis, \np{355} (1991) 649;
\np{329} (1990) 27;
I. Antoniadis, K. Narain, and T. Taylor, \pl{267} (1991) 37;
V. Kaplunovsky and J. Louis, \np{422} (1994) 57;
J. Casas, Z. Lalak, C. Munoz and G. Ross, \np{347} (1993) 22;
T. Banks, D. Kaplan, and A. Nelson, \prd{49} (1994) 779;
B. de Carlos,
J. Casas, F. Quevedo and E. Roulet, \pl{318} (1993) 447.}
\lref\kap{V. Kaplunovsky, \np{307} (1988) 145; \np{382} (1992) 436.}
\lref\dkv{L. Dixon, V. Kaplunovsky and C. Vafa, \np{294} (1987) 43.}
\lref\sy{A. Schellekens and S. Yankielowicz, \np{327} (1989) 673; {\it ibid.}
Int. Jour. Mod. Phys. A5, (1990) 2903.}
\lref\sgen{W. Lerche, A. Schellekens and N. Warner, Phys. Rep. 177 (1989) 1.}
\lref\schel{A. Schellekens, \pl{237} (1990) 363.}
\lref\ki{K. Intrilligator, \np{332} (1990) 541.}
\lref\ginsparg{P. Ginsparg, {\it Applied Conformal Field Theory}, Les Houches
lectures (1988).}
\lref\ver{E. Verlinde, \np{300} [FS22] (1988) 360.}
\lref\ckt{ S. Chaudhuri, H. Kawai and S.-H. H. Tye,
\np{322} (1989) 373.}
\lref\slansky{R. Slansky, Phys. Rep. 79 (1981) 1.}
\lref\bdfm{T. Banks, L. Dixon, D. Friedan and E. Martinec,
\np{299} (1988) 613, and references therein.}
\lref\ds{M. Dine and N. Seiberg, \pl{162} (1985) 65; \np{301} (1988) 357.}
\lref\ghmr{D. Gross, J. Harvey, E. Martinec and R. Rohm, \np{256} (1985) 253.}
\lref\bdix{T. Banks and L. Dixon, \np{307} (1988) 93.}
\lref\fms{D. Friedan, E. Martinec and S. Shenker, \np{271} (1986) 93.}
\lref\gsw{M. Green, J. Schwarz and E. Witten (1986).}
\lref\dfms{L. Dixon, D. Friedan, E. Martinec, and S. Shenker,
\np{282} (1987) 13.}
\lref\nao{S. Hamidi and C. Vafa, \np{279} (1987) 465;  D. Freed and
C. Vafa, \cmp{110} 349 (1987).}
\lref\aso{K. Narain, M. Sarmadi and C. Vafa, \np{288} (1987) 551.}
\lref\dhvw{L. Dixon, J. Harvey, C. Vafa, and E. Witten, \np{261} (1985)
678; \np{274} (1986) 285.}
\lref\gw{D. Gepner and E. Witten, \np{278} (1986) 493.}
\lref\gq{V. A. Fateev and A. B. Zamolodchikov, JETP 66 (1985) 215; {\it ibid.}
JETP 63 913; D. Gepner and Z. Qiu, \np{285} (1987) 423.}
\lref\gep{D. Gepner, \np{287} (1987) 111; {\it ibid.}, \np{296} (1987) 757.}
\lref\narain{K. S. Narain, \pl{169} (1986) 41; K. S. Narain, M. H. Sarmadi
and E. Witten, \np{279} (1987) 369.}
\lref\dis{J. Distler and S. Kachru, \np{413} (1994) 213.}
\lref\strom{A. Strominger and E. Witten, \cmp{101} (1985) 341.}
\lref\wit{E. Witten, \np{403} (1993) 159; {\it ibid.} Int. Jour. Mod. Phys.
A9 (1994) 4783.}
\lref\bd{T. Banks and M. Dine, {\it Coping with strongly coupled
string theory}, Rutgers preprint RU-50-94, hep-th9406132.}
\lref\go{P. Goddard and D. Olive, Int. Jour. Mod. Phys. {\bf A1}, (1986) 303.}
\lref\gko{P. Goddard, D. Kent and D. Olive, \pl{152}, (1985) 88.}
\lref\gno{P. Goddard, W. Nahm and D. Olive, \pl{160} (1985) 111.}
\lref\tye{Henry Tye, private communication.}
\lref\font{L. Ibanez, J. Mas, H-P Nilles, and F. Quevedo,
\np{301} (1988) 157.}
\lref\can{See, for example, B. Greene,
{\it Lectures on quantum geometry},
Lectures at the Jerusalem Winter
School, 1994, and references therein.}
\lref\feng{Jonathan Feng, private communication.}
\lref\agv{L. Alvarez-Gaume, P. Ginsparg, G. Moore, and C. Vafa,
\pl{171} (1986) 155.}
\lref\kltt{H. Kawai, D. Lewellen, and S.-H. H. Tye, \prd{34} (1986) 3794.}
\lref\flip{I. Antoniadis, J. Ellis, J. Hagelin, and D. Nanopoulos,
\pl{208}  (1988) 209; See also J. Lopez, D. Nanopoulos, and
A. Zichichi, hep-ph/9502414.}
\lref\cleaver{G. Cleaver, {\it GUTs with adjoint Higgs fields from
superstrings},  Ohio State preprint OHSTPY-HEP-T-94-007, hep-th/9409096.}
\lref\nonwitt{E. Witten, \cmp{92} (1984) 455.}
\lref\god{P. Goddard, W. Nahm, D. Olive, and A. Schwimmer, \cmp{107}
(1986) 179; D. Bernard and J. Thierry-Mieg, \cmp{111} (1987) 181;
for an introduction to free fermions and current algebras, see
T. Banks, {\it Lectures on conformal field theory}, Proceedings of
the 1987 TASI school, R. Slansky and G. West eds., World Scientific,
Singapore, 1988.}
\lref\horvath{Z. Horvath and L. Palla, \pl{229} (1989) 368.}
%%%%%%%%%%%%%%%%%%%%%%%%%%%%%%%%%%%%%%%%%%%%%

\newsec{Introduction}

It is important in string theory to develop the dictionary that
translates between four dimensional spacetime physics and the
world-sheet properties of the string vacuum \bdfm \ds . This will
enable us to understand how specific
phenomenological properties of possible interest in the low energy
effective field theory are realized in superstring
unification \ss \gs \st . Much of the work to date
in superstring phenomenology has focussed on the
$(N_R,N_L)$$=$$(2,2)$
compactifications \chsw\ of the ten-dimensional
$E_8$$\times$$E_8^{'}$ heterotic superstring \ghmr .
The larger class of $(2,0)$ vacua \dix\
has, however, remained largely
unexplored except for the simplest abelian orbifold
compactifications \dhvw \aso \font , a subset of which have an
equivalent free fermionic
realization \klt \abk .

In recent work \us , we have used {\it real
fermionization}\myfoot{We use the expression ``real fermionization''
to distinguish this approach from
free fermionic formulations \klt \abk \kln\
which assume a realization of the internal conformal field theory
in either Weyl or Ising fermions, but have no {\it unpaired} Majorana-Weyl
fermions.} \klst \dhl \joel\ to understand how specific features of
interest in the massless spectrum and tree-level couplings of
the low-energy effective field theory are realized in exact
solutions to string theory. Our starting point is the low-energy
effective field theory. We will apply our knowledge of conformal field
theory to find consistent ground states of string theory
which embed spacetime features of possible phenomenological interest.
Our preliminary results suggest many intriguing possibilities for
phenomenology that are not present in either the $(2,2)$
solutions or the known $(2,0)$ orbifold compactifications.
Some preliminary results have also been obtained by G. Cleaver \cleaver .
L. Ibanez and collaborators \ibanez\ have recently begun a
similar study of the phenomenological implications of
higher level current algebras within the orbifold
construction.

One of our goals is to make contact between string theory and
more conventional field theoretic unification models.
There are many indications that such a cross-fertilization
of ideas would be fruitful. In the coming years the detailed
exploration of the electroweak scale and the neutrino sector
is likely to yield additional clues about short-distance physics
besides the preliminary evidence for gauge coupling unification.
In addition, increasingly accurate determinations of the parameters
of the Standard Model will provide tighter constraints on
unification schemes.
The motivation
for string theory is rooted in the successful unification of parity
violating gauge interactions, quantum mechanics, and gravity \ss \gs \st .
It is therefore important to establish to what extent the low-energy
particle physics consequences of string theory are robust.
The string consistency conditions of modular invariance and
world-sheet supersymmetry are extremely restrictive
constraints on the spectrum.
Thus we may expect guidance and insights for
unification model builders by requiring string consistency of
the effective field theory at the unification scale.

Supersymmetric grand unification models \guts\ suggest a picture in which
radiative electroweak symmetry breaking and the large top quark
mass are generated from a GUT-scale effective superpotential with
a single third generation Yukawa coupling \rad .
The distinct hierarchies in the pattern of fermion masses and mixings
at the electroweak scale may be generated, in part,
by higher dimension operators in the effective superpotential  \textures .
The recent results of Anderson, Dimopoulos, Hall, Raby, and Starkman
\adhrs\ illustrate that the presence of a {\it small} number of
higher dimension operators in the GUT-scale effective superpotential
may be adequate to generate the observed masses and mixings.
These higher dimension operators \adhrs\ are suppressed by powers of $M_{G}$
over $M_X$, where $M_{G}$$\approx$$10^{16}$ GeV, and $M_X$ is
another superheavy mass scale $\approx$$10^{17}$ GeV. Restrictive
flavor-sensitive selection rules are required in such scenarios to
eliminate unwanted higher dimension operators and Yukawa couplings
from the superpotential. Even more restrictive selection rules
will be necessary in order to generate GUT scale masses for
the triplet Higgs fields while keeping the supersymmetric
Standard Model Higgs fields light \dw \babu . Such restrictive
symmetries appear unnatural from the point of view of an effective field
theory.
It is
well-known in a general sense that string theory can provide such
selection rules \dine . Less well-known is the possibility of
using real fermionization to produce models
which resemble
conventional supersymmetric GUTs \dhl \fer \joel .
Real fermionization
is also relevant to recent ideas about supersymmetric
textures which do {\it not }invoke GUTs \rametal .

Finding explicit solutions to string theory that realize the
required massless spectrum and selection rules of such
mass matrix models will both provide guidance to model builders \moha\
and eventually give deeper insight into the origin of fermion masses
and mixings. It should be noted that {\it unification} in the context of
superstring theory has broader significance than the unification
of the gauge couplings and (or) Yukawa couplings. The
dynamical supersymmetry breaking sector, and a mechanism for feeding
supersymmetry breaking to the low-energy matter, must be built
into any consistent solution to string theory. Thus,
string consistency is a
powerful guiding principle in building {\it complete}
supersymmetric models, which do not merely parametrize the
weak scale effective Lagrangian but which also specify
the origin of the soft supersymmetry
breaking parameters.

Free fermionization is one of the oldest techniques known to string
theorists and is the basis for the Ramond-Neveu-Schwarz formulation
of the superstring \rns \gso \st . The use of generalized GSO projections
\gso\ to construct new solutions to string theory,
given a consistent solution,
was introduced in the context of the ten dimensional heterotic superstring
in \sw \agv \kltt . The ten dimensional ground states include a
(non-supersymmetric) solution where the gauge symmetry is realized at
level two \kltt . In \klt \abk\ this
technique was applied to construct ground states with four dimensional
Lorentz invariance. The fermionic formulation is based on the notion of
current algebras and free fermionic
representation theory \rns \nonwitt \gno .
A comprehensive discussion of non-renormalizable tree-level superpotential
couplings can be found in \kln . Methods for analysing moduli
dependence are given in \cs \joel \lnyb , but
these require further development.

A number of models
of phenomenological interest have been constructed
using free fermionization \flip \lnya \far . These models
contain three generations of light chiral fermions and
gauge groups like $SU(3)$$\times$$SU(2)$$\times$$U(1)$ or
``flipped'' $SU(5)$, realized by Weyl
fermions on the world-sheet as current algebras
at level one. The superpotential of the resulting low energy effective
field theory has been computed for these models, using the techniques
described in \kln . One then discovers interesting flavor-sensitive
selection rules which restrict Yukawa couplings.

These realistic models belong to the subclass of free fermionic
models which contain Weyl and Ising fermions, but do not
contain any unpaired Majorana-Weyl fermions,
which we call {\it real}
fermions.\myfoot{Properly speaking Ising fermions, which are
right-left pairs of Majorana-Weyl fermions, are also real
fermions. However it is very convenient for our analysis to
let ``real'' denote only unpaired fermions, and identify Ising
fermions separately. We hope that this usage does not cause
confusion with respect to references \kln\ and \klst , where
``real fermions'' includes Ising fermions.}
Models with only Weyl fermions produce simply-laced current algebras
with Kac-Moody level one. This is because the local algebra of
$n$ Weyl fermions has central charge $n$ and always contains $n$
abelian currents.
Models with both Weyl and Ising fermions can have reduced rank,
because the Ising fermions soak up central
charge without producing abelian currents. This also allows
realizations of $SO(2n$$+$$1)$ at level one \god ,
and $SU(2)$ at level two \dhl .

Local algebras of $2n$ real fermions have central charge $n$
and some number of abelian currents which is {\it variable}
between zero and $n$. This richer set of local algebras allows us
to realize current algebras which cannot be obtained in the
subclass of models just discussed. In particular,
real fermionization enables us to realize many current algebras at
higher level. This in turn allows the appearance of adjoint
Higgs in the massless spectrum, as needed for conventional GUT's \dhl .
Real fermionization also provides new embeddings of level one
current algebras, and new possibilities for discrete symmetries
in the effective field theory.
We thus aim to exploit the techniques and successes of
\flip \lnya \far \kln\ while exploring a more general construction.

A non-trivial extension of these techniques is required when the underlying
conformal field theory includes real fermions.
The source of the difficulty is phase ambiguities in the explicit
definition of the GSO projections and higher loop modular
transformations for the real fermion conformal field
theory. These phases play a crucial role in determining
the massless spectrum and tree level
couplings of the resulting models.
A first attempt at
resolving these ambiguities was made in \klst . We supplement
that analysis by developing additional {\it tree-level}
checks for string consistency.

The outline of this paper is as follows. In section 2 we review the
well-known correspondence between gauge symmetry in spacetime and
current algebras on the world-sheet \ghmr . This introduces the
notion of world-sheet constraint algebras underlying the properties
of the low energy effective field theory. In section 3 we explain
in general how
a higher level current algebra can be realized in the tensor product of
{\it constituent} conformal field theories.
We illustrate this with a toy model.
Free fermion conformal field theories that
embed both the gauge bosons and the chiral superfields transforming
under such a current algebra, can be built
into a consistent solution to string theory by using the real
fermionization prescription of \klst .
We explain how this works in the pedagogic
discussion in section 4, presenting two examples
with distinct fermionic embeddings of $SO(10)$.
All of the results in this section were
obtained with the use of a symbolic manipulation package developed
by us \docspec . In section 5 we address some of the formal
developments necessary to understand real fermionization at a
more fundamental level than the prescription of \klst .
We use Verlinde's theorem \ver\ to relate the
tree-level fusion algebra to the one-loop spin structure blocks
in a way which allows unambiguous computation of
the tree level correlators for real fermions. Combined with the
methods of, e.g., \kln , this
will enable us to eventually automate the extraction of the tree-level
superpotential. Our better understanding of real fermionization
also allows us to probe cases of real fermion spin structures
where the prescription of \klst\ breaks down.
In the conclusion we make a critical appraisal of free
fermionization, list some remaining problems,
and discuss extensions of our methodology.
We do not attempt to display any phenomenologically compelling
models in this paper.

\newsec{Spacetime symmetries and world-sheet operator algebras}

The two-dimensional gauge principle of heterotic string theory
is $(1,0)$ superconformal invariance \ghmr \st . In
light-cone gauge,\myfoot{We restrict ourselves to
spacetime backgrounds with four dimensional Lorentz invariance.}
the decoupling of timelike and longitudinal degrees
of freedom results in a unitary conformal field theory,
with a Hilbert space of positive norm. The field content
includes the non-compact transverse spacetime coordinates,
$X^{\mu}$$=$${\bar X}^{\mu}
({\bar z})$$+$$X^{\mu}(z)$,  $\mu$$=$$1$,$2$,  and
their Majorana-Weyl fermion superpartners,  $\psi^{\mu}({\bar z})$.
In addition, there is an internal $(1,0)$ unitary conformal
field theory of central charge $(9,22)$. Every physical state
corresponds to the lower component
of a conformal dimension $(h_R,h_L)$$=$$({1\over 2},1)$ {\it world-sheet}
superfield
transforming under the $(1,0)$ superconformal constraint algebra.

The notion of finding world-sheet constraint algebras
related to spacetime properties of the low-energy
effective field theory was first explored in
references \bdfm \bdix . We begin by reviewing the familiar
example of gauge symmetry in order to
explain how the constraint algebra can be used to build a
solution to string theory embedding a specific low energy
spectrum of fields.

In an $N$$=$$1$ spacetime supersymmetric vacuum all of the
gauge symmetries are associated with the left-moving conformal field
theory \ghmr . Then there must exist vertex
operators of conformal dimension $({1\over 2},1)$ which transform
as spacetime vectors, corresponding to gauge bosons:
\eqn\gbos{
V^a (z , \zbar   ) ~=~ \zeta^{\mu} \psi^{\mu} (\zbar) J^a(z) e^{i k \cdot X}
{}~~~~~~,}
\noindent where $\zeta^{\mu}$ is the transverse polarization vector,
$ \zeta \cdot k $$=$$k \cdot k $$=$$0$, and $J^a(z)$ is a dimension $(0,1)$
primary
field in the left-moving internal conformal field theory. Gauge symmetry
is therefore a consequence of
an extension of the $(1,0)$ superconformal constraint algebra by
dimension $(0,1)$ currents. The presence of the gauge bosons in the
spectrum of massless fields implies that any chiral superfields that
appear in the spectrum
must satisfy the selection rules imposed by gauge invariance.
In world-sheet language this implies strict agreement with the
fusion rules of the world-sheet current algebra.

The operator product algebra of the dimension $(0,1)$
operators, $J^a (z)$, determines the structure constants and
Schwinger term of a current
algebra\myfoot{We will
use the term current algebra for what is often referred
to as an affine Kac-Moody algebra \kac \go . We
will assume that the low energy gauge symmetry is related to a
compact Lie group.}:
\eqn\kma{
J^a (z) J^b(w) ~~=~~ {{\delta^{ab}(k\psi^2/2)}\over{(z-w)^2}}
{}~+~ {{i f^{abc} J_c}\over{(z-w)}}
{}~+~ \cdots ~~~~.
}
\noindent where $\psi^2$ is the length-squared of the highest root.
This current algebra is, in general, based on the
product of simple non-abelian and abelian group factors.
The central charge from any simple group
factor is given by the formula
\eqn\central{
c_k (G) ~~=~~ {{k ~ dim(G)}\over{ k+\tilde{h}}} ~~~~.
}
\noindent The dual Coxeter number, $\tilde{h}$, is
equal to $C_A/\psi^2$, where
$C_A$ is the quadratic Casimir of the adjoint
representation.
The {\it Kac-Moody level}, $k$,
is restricted to take integer values due to the
unitarity of the conformal field theory.
It is common to normalize $\psi^2$ to 2 (or 1) so that the level
coincides with (or is twice) the coefficient of the double
pole term in \kma . For our purposes it is more natural to
normalize the coefficient of the double pole term to 1; the level
is then read off from the norm of the roots.

In order to build a solution containing
a specific low-energy spectrum of vector and chiral
superfields, it suffices to find a realization of those gauge bosons
which correspond to the {\it simple roots}, and the chiral superfields
corresponding to the {\it highest weights} of the desired irreducible
representations. The current algebra will automatically generate
complete supermultiplets in the solution {\it if} care is taken to preserve
the string consistency conditions of world sheet supersymmetry and
modular invariance.

Thus, Lorentz invariance, spacetime supersymmetry
and gauge invariance determine, in part, the emission vertex of any
chiral superfield. Consider, for example, the vertex operator
associated with
a fixed helicity of a chiral superfield transforming as a
spacetime fermion, $V^{+}_r (z , \zbar )$. The vertex operator
corresponding to the highest weight of an irreducible representation $r$
will take the form,
\eqn\matter{
V^+_r (z , \zbar ) ~=~ S( \zbar )  O( \zbar ) f_r (z) F(z) e^{i  k \cdot X}
{}~~~~~.
}
\noindent We have left unspecified the dimension $({3 \over 8},0)$
primary field, $O( \zbar )$, which must occur in the Ramond sector of
the internal superconformal field theory; its form is restricted
by the spacetime supersymmetry currents.
$S( \zbar )$, is a dimension $({1\over 8},0)$ spin
field in the Ramond sector of the conformal field theory of the Majorana-Weyl
fermions $\psi^{\mu}( \zbar )$.
The Kac-Moody primary field $f_r (z)$ is of dimension $(0,h_r )$,
and $F(z)$ is a gauge singlet of dimension $(0,1- h_r )$.

With higher level
realizations of the current algebra,
new matter representations can appear consistent with the requirement
of unitarity of the underlying conformal field theory.
This introduces new options for spacetime gauge and
gravitational anomaly cancellation,
depending on which chiral fermion representations appear
in the massless spectrum.
A detailed tabulation of which representations and
conformal dimensions are allowed in an affine Lie
algebra at arbitrary level can be found
in \patera\ and \go . We should emphasize that, while
unitarity is a restriction on which representations
can appear at any given level, not every allowed
representation need appear in a conformal field theory described by
an asymmetric modular invariant.

\newsec{Embedding higher level current algebras}

The easiest way to realize a specific spacetime gauge symmetry
in a consistent solution to string theory is to find an
embedding of the current algebra in the tensor product of
simple {\it constituent} conformal field theories. The best
known constituents are free bosons and free fermions. However,
as will become apparent, the method can be applied more
generally.

The basic idea underlying the higher level current
algebra realization is very simple. We begin by
realizing the $r$ abelian currents of the Cartan subalgebra
of the group in a conformal field theory denoted
as $CFT_{A}$. An abelian generator can always
be realized by a chiral boson {\it with no loss of generality.}
If we are realizing a non-abelian current algebra
the chiral bosons have rational conformal
dimensions (see, for example, \go \sgen ). Thus $CFT_{A}$ is
constructed using $r$ chiral bosons with conformal dimensions,
$h_L$$=$$p_L^2/2$$=$${m/n}$,
with $m$, $n$ integers.

For a higher level realization it is not possible to
construct the remaining currents of the non-abelian
current algebra using only operators of the free boson
conformal field theory, $CFT_{A}$.\myfoot{Higher level
realizations using {\it twisted} free bosons are possible:
see \horvath .}
Thus what we actually
need is a tensor product of $CFT_{A}$ with some other
constituents, which we will denote collectively as $CFT_B$.
In this paper we restrict ourselves to the cases
where $CFT_B$ is constructed using unpaired Majorana-Weyl
(real) fermions.
This is a {\it strong} restriction on which gauge
groups and representations can be obtained in this
class of solutions. The obvious generalization
is to allow as constituents of $CFT_B$ any of the
unitary conformal field theories with central charge
$c$$<$$1$ \fqs . These conformal field theories have a
finite number of chiral primaries under the Virasoro
algebra and rational conformal dimensions, $h_i$$<$$1$.
They have no spin one currents.
The corresponding Virasoro characters, which enter the
string partition function, have well-defined
modular transformation
properties.

If the tensor product $CFT_A$$\times$$CFT_B$ successfully
realizes a current algebra, then the total central
charge $c_A$$+$$c_B$ must {\it at least} equal $c_k(G)$.
If $c_A$$+$$c_B$$>$$c_k(G)$ this implies that we have realized,
in addition to the higher level current algebra,
some {\it other} holomorphic algebra which contains
no currents. We will refer
to this other algebra
as a {\it discrete holomorphic operator algebra}.

Thus the (left-moving) stress tensor for a higher level current algebra
realization has, in general, two distinct decompositions:
\eqn\stress{
\eqalign{
T &= T_A + T_B \cr
&= T_{KM} + T_{discrete}\qquad ,
}}
where $T_{KM}$ denotes the Sugawara form of the stress
tensor of the higher level current algebra, and
$T_{discrete}$ denotes the coset algebra formally
defined by the relation \stress .

Two observations of considerable practical
importance are as follows. The rank of the
low-energy gauge symmetry in a four dimensional
ground state is bounded by the central charge
of the left-moving internal conformal field theory,
$\sum_i rank(G_i)$ $\le$ $22$. Also,
the dimensions of individual matter representations
that can appear at the massless level are bounded by
the condition, $\sum_i h^i_L $ $\le$ $1$ \dkv \font .

The conformal field theory of a chiral boson, $\phi(z)$, with
rational-valued momentum, $p$, is equivalent to that of
a Weyl fermion, $\lambda(z)$, with fermionic charge, $Q$:
\eqn\compl{\eqalign{
\partial \phi ~ &\to ~ : \lambda^{\dagger} \lambda : \cr
{\hat p} ~ &\to ~ { \widehat Q} ~=~ {\widehat F} ~-~ {v \over n}
{\bf 1} ~~~~~.
}}
\noindent Here ${\widehat F}$ is the fermion
number operator, and the vacuum fermionic charge,
$v/n$, is rational-valued. The abelian
current is realized by the Weyl fermion bilinear.
Fermionic representations of current algebras that
utilize fermion bilinears are well-known. The
non-simply-laced algebras at level one can be
realized by Majorana-Weyl fermions. For example,
the generators of $SO(2n+1)$ are realized by
$n$ Weyl fermions and a single Majorana-Weyl fermion, or
equivalently, $2n+1$ Majorana-Weyl fermions \god .
The currents are the $2n(2n+1)/2$ Majorana-Weyl
fermion bilinear pairs.

When we realize the Cartan currents using Weyl
fermion bilinears, every distinct group weight will
be realized as a unique set of fermionic charges.
This representation of weights in a basis defined by
fermionic charges is fixed once we specify the fermionic
charges of the $r$ simple roots \slansky .
We then identify in $CFT_A$ holomorphic operators,
$\phi^a_{q_1,\ldots q_r}(z)$
with the correct fermionic charges $(q_1,\ldots q_r)$
to represent all
of the currents, $J^a(z)$, of the higher level algebra.
Since these primaries may not have conformal dimension 1,
we then must identify other operators in $CFT_B$ to
make up the difference. Thus
\eqn\curdecomp{
J^a(z) = \phi^a_{q_1,\cdots q_r}(z)\times \phi^a_B(z) \qquad .
}
The above also holds for chiral bosons when we map weights into momenta.

\subsec{Canonical Embeddings}

Let us explain, from first principles,
how one can identify a realization of some given current algebra at
arbitrary level, assuming explicit knowledge of the conformal
dimensions, operator product coefficients, and Virasoro characters
of the chiral primaries of the constituent conformal field
theories.

There are many possible free field embeddings of any given current algebra.
We will refer to the embedding with the lowest possible total conformal
anomaly as the {\it canonical} embedding.
One advantage of using a canonical
embedding of the roots (e.g., the standard Cartan-Weyl basis for
a level one realization) is that the model builder avoids
the pitfall of unexpected extra gauge symmetry
such as U(1) factors in the final solution.

We begin with a realization of the Cartan subalgebra
of the group. Each of the $r$ abelian currents is
realized by a chiral boson
\eqn\abelian{
h_i ~=~ \partial \phi_i ~~~~~~~~i~=~1, \cdots , ~r ~~~,
}
\noindent where $r$ is the rank of the gauge group.
These are operators of conformal dimension one. Let
us assume that the momenta of the individual chiral
bosons are quantized such that
\eqn\momenta{
\phi_i(\sigma_1 + 2 \pi , \sigma_2) ~=~
\phi_i(\sigma_1, \sigma_2) ~+~ 2 \pi p_i ~~~~.
}
\noindent Consider vertex operators of non-zero momentum
\eqn\fkc{
V^{(\pm)}_j ~=~ C_j({\hat p}) : e^{\pm i{\bf p}_j \cdot {\bf \phi}} :
{}~~~,
}
\noindent where ${\bf p}_j$ and ${\bf \phi}$ are $r$ dimensional
vectors, and the $C_j({\hat p})$ are cocycle operators. This is
the familiar vertex operator construction used in the
$E_8$$\times$$E_8$ heterotic string \st : if the ${\bf p}_j $ lie
on the root-lattice of a simply-laced group the commutation
relations of the vertex operators, with cocycle operators
appropriately defined, will reproduce the
structure constants of the associated current algebra.
The normalization of the abelian currents is not fixed until
we specify the realization of the nonzero roots.

Now consider a specific example of this construction in the
context of heterotic string theory. Begin with five copies of the
root lattice of $SU(2)$
\eqn\su{
([ \pm \sqrt 2 , 0,0,0,0]) \qquad ,
}
\noindent where the square brackets denote permutations, and
we have normalized the roots to length $\alpha^2$$=$$2$.
Let us assume that this lattice is embedded
in the $22$ dimensional sublattice of an even
self-dual Lorentzian lattice of dimension
$(6,22)$ \narain \st . The states corresponding to the roots of $(SU(2))^5$
given in
\su\ will then appear at the massless level, with $p_L^2$$=$$2$, $h_L$$=1$, and
correspond to spacetime gauge bosons. The realization of the gauge symmetry
is at level one. From the properties of self-dual lattices, it
follows that the weight lattices of $(SU(2))^5$
\eqn\ssu{
([\pmsq , 0,0,0,0])
}
\noindent are present in the $(6,22)$ dimensional lattice \sgen .
Ignoring the precise constraints from modular invariance, imagine
that we perform a sequence of orbifold twists accompanied by shift vectors
embedded in the $(SU(2))^5$ lattice whose net effect is to project
out the {\it individual} roots and weights but leave intact
the lattice points
\eqn\sus{
([\pmsq , \pmsq , 0,0,0]) ~~~~~,
}
\noindent where all permutations are included. The {\it counting}
of states is correct to fill out the adjoint representation
of the group $SO(10)$, $5 \cdot 4 \cdot 2 ~+~5$ giving a total of $45$
states if we include the states corresponding to the
five abelian currents.

Suppose we rescale the normalization of the abelian currents
by a factor of two. Then the length of the lattice vectors in
\sus\ is exactly what is needed for a level two realization of the
gauge symmetry. The only problem is that the states of
non-zero momentum no longer appear at the {\it massless} level
because the (left) conformal dimension is only
${1\over 2} \cdot p_L^2$$=$${1\over 2}$. This problem is easily fixed.
The central charge of $SO(10)$ at level $2$ can be read off
from the formula \central\ given in the previous section,
where $C_A$$=$$2(2n-2)$ for $SO(2n)$. The central charge of the
embedding conformal field theory of five chiral bosons is $c$$=$$5$.
Thus, if we can find a (rational) conformal field theory with central charge
$c$$>$$4$, primary fields of conformal dimension ${1\over 2}$, and
{\it no} dimension one currents, by tensoring together the
two conformal field theories it should be
possible to find an embedding of these states at the massless level.
A necessary requirement is that we exactly match the conformal
dimensions and counting of states given above {\it without}
modifying their fusion rules.

Let us outline how to find such an embedding for our
toy model.\myfoot{The reader will recognize
an obvious parallel with the asymmetric orbifold
construction in the discussion that follows.}
The first five left-moving entries of the $(6,22)$
dimensional lattice before twisting have already been
determined \su , \ssu . Let us assume that the next eight
entries embed the root-lattice of $SO(16)$
\eqn\sosix{
([\pm 1, \pm 1, 0,0,0,0,0,0])
{}~~~~.
}
\noindent Together with the spinor and conjugate spinor weights
of $SO(16)$,
\eqn\sowe{
(\pm\ha ,\pm\ha ,\pm\ha ,\pm\ha ,\pm\ha ,\pm\ha ,\pm\ha ,\pm\ha )
{}~~~~,
}
\noindent one obtains the $E_8$ lattice. This lattice is easily
embedded in an even self-dual Lorentzian lattice given by the sum of the
root and weight lattices of $(SU(2)_L)^6 \times (SU(2)_R)^6 \times
E_8 \times E_8' $ \narain . The self-dual lattice describes the
compactification
of the ten dimensional $E_8$$\times$$E_8'$ heterotic string on an
$(SU(2))^6$ torus.

The conformal field theory underlying the $E_8$ lattice
has a fermionic
representation \ghmr \st . The eight chiral bosons can be
fermionized as follows:
\eqn\bosoniz{
\eqalign{
\partial \phi_i ~&\to~ : \lambda_i^{\dagger} \lambda_i : ~~~~~
i=7, \cdots , 14 ~~~ \cr
e^{\phi_i} ~&\to~ (-1)^{\hat F_i} \lambda_i \cr
{\hat p}_i ~ &\to ~ {\widehat F_i} ~-~ {{v_i}\over 2} {\bf 1} \qquad .
}
}
\noindent The equivalence between momentum and fermionic
charge for momentum quantized in half-integer units,
$p_i = n/2$, implies that the
conformal field theory of the Weyl fermions has two sectors.
The two sectors correspond to choosing Neveu-Schwarz
(antiperiodic) or Ramond (periodic) boundary conditions
for the fermions, respectively, $v_i$$=$$0,1$:
\eqn\bound{
\lambda_i(\sigma_1 + 2 \pi, \sigma_2) ~=~
- e^{\pi i v_i } \lambda_i(\sigma_1, \sigma_2) ~~~~~.
}
\noindent The roots of $SO(16)$ correspond to oscillator
excitations in the Neveu-Schwarz sector. The spinor weights
given in \sowe\ correspond to states in the Ramond sector,
with $F_i$$=$$0,1$, and $v_i$$=$$1$, for all $i$.
In the one-loop vacuum amplitude this sector is
labelled by a vector specifying the boundary conditions
of the individual fermions, $v_i, ~i=1, \cdots , 8$,
\eqn\disccomp{
 (1~1~1~1~1~1~1~1) ~~~~.
}
\noindent Thus, in the absence of constraints from any
other sectors, this sector contributes the $2^8$
spinor and conjugate spinor weights of $SO(16)$
in the one-loop
vacuum amplitude.

For convenience, we can rewrite the Weyl (complex) fermions
as Majorana-Weyl fermions,
$ \lambda_i $$=$$\psi_i^{(1)} +i \psi_i^{(2)}$.
The two Majorana-Weyl fermions associated with each of the eight Weyl
fermions share the same boundary condition in every sector
summed over in the one-loop vacuum amplitude. Implicitly,
we are now allowing for the possibility of
Majorana-Weyl fermions which are no longer paired into complex fermions.
Some of these may be right-left paired into Majorana (Ising) fermions.
Any Majorana-Weyl fermions which are truely unpaired we call
{\it real} fermions.
In the absence of a complexification
of the Majorana-Weyl fermions, a conserved fermionic charge,
or equivalently, a
conserved bosonic momentum, can no longer be defined.
We can re-label the sector \disccomp\ contributing
the spinor weights of $SO(16)$
by the corresponding boundary condition vector
($v_i$$=$$1$, $i=1, \cdots 16$) for {\it sixteen} real fermions:
\eqn\discreal{
(11~11~11~11~11~11~11~11) ~~~~.
}

Ignoring once again the constraints from modular invariance,
consider the possibility of blocks of {\it chiral} $Z_2$ twists on the
$E_8$ lattice accompanied by the shift vectors embedded in the
$(SU(2))^5$ lattice such that all of the $E_8$ gauge symmetry is
broken to a discrete subgroup.
This corresponds to introducing new sectors in the one-loop vacuum
amplitude which contribute states of non-zero momentum in the
conformal field theory of the chiral bosons,
$\phi_i$, $i$$=$$1$, $\cdots$, $5$,
corresponding to the lattice points \sus , matched with the tensor product of
Ramond ground states for blocks of eight real fermions chosen from
the set, $\psi_i^{(j)}$, $i$$=$$1$, $\cdots $, $8$, and $j=1,2$. In order to
break all of the $E_8$ gauge symmetry we need to include at least
four sectors in the one-loop vacuum amplitude, corresponding to the
following boundary condition vectors for the sixteen real fermions:
\eqn\twists{
\eqalign{
 &(1111 ~1111 ~0000 ~0000) \cr
 &(0000 ~1111 ~1111 ~0000) \cr
 &(1100 ~1100 ~1100 ~1100) \cr
 &(1010 ~1010 ~1010 ~1010)
{}~~~~.}
}
\noindent The contribution to the left conformal dimension
from the Ramond vacuum energy in each of these sectors is
${1\over 16} \cdot 8 $$=$${1\over 2}$.
Therefore, oscillator excitations described
by fermion bilinears of the form, $ : \psi_j \psi_k : $, contribute with
conformal dimension {\it greater} than one in these sectors and are pushed
up to the massive level. The sectors \twists\ also act as constraints
on the {\it untwisted} sector, i.e., the sector with all fermions
in the Neveu-Schwarz vacuum, so that these dimension one states are
projected out of the spectrum by the requirement of modular invariance. Thus
the untwisted sector does not contain any currents.
Of course, one must still be concerned with additional dimension one
states that can contribute from twisted sectors.
Choosing the projections on the
spectrum such that no additional dimension one currents appear requires a
detailed knowledge of the constraints from one-loop modular invariance.
While this certainly could be done, we will not pursue this toy
model any further. Certain elements of the toy model can, however, be
recognized in the examples of section 4.

The embedding \sus\ of the roots of $SO(10)$ in the doublets of five
copies of $SU(2)$ is a special case of the embedding of the roots of
$SO(2n)$ at level $k$$=$$2$ in the fundamental weight-lattices of the group
$(SU(2))^n$. The pattern further generalizes to an embedding of the
roots of $SO(2n)$ at level $k$ in the momentum lattice of $n$ chiral bosons,
with momentum quantized in units of $1$$/$$\sqrt{k}$.
Embeddings of
the roots of the special unitary groups can be worked out by the same method.

\subsec{Fermionic Embeddings}

Now let us specialize to the case where the $c$$=$$1$ constituents
of $CFT_A$ are Weyl fermions and the constituents of $CFT_B$
are $c$$=$${1\over 2}$ Majorana-Weyl fermions.

It is important to distinguish
between a fermionic {\it embedding}
and a fermionic {\it representation}
of a current algebra.
A fermionic embedding is simply
a mapping of the roots of a Lie algebra into fermionic charges.
A fermionic representation is an embedding where
the total conformal anomaly of the fermions
equals the central charge
of the Kac-Moody algebra.
An example of a higher level fermionic representation is
$SU(2)$ at level two realized by three Majorana-Weyl fermions.

Fermionic representations may or may not exist
depending on the group and
the level of the current algebra.
The orthogonal groups at level
one have fermionic representations. But the
special unitary groups at level one are only obtained in the
fermionic embedding of the group $SU(n)$$\times$$U(1)$.
The `extra' U(1) in a fermionic embedding cannot be broken by
standard stringy symmetry breaking techniques, e.g., a
$Z_2$ twist, without simultaneously breaking the nonabelian symmetry.

These statements have counterparts for fermionic realizations
of higher level current algebras.
A fermionic embedding determines the level of the
current algebra by fixing
the lengths-squared of the nonzero roots.
To be precise, let
$\overrightarrow{Q}$$=$$(q_1,q_2,\ldots q_n)$
denote the fermionic charges of a root; then $\overrightarrow{Q}{}^2$
must have the same value for all the roots (all the long
roots if the group is not simply laced). The level is then given by \dhl :
\eqn\levelis{
k = {2\over \overrightarrow{Q}{}^2} \qquad .
}
An example of a higher level fermionic embedding is
the minimal {\it fermionic} embedding of the
roots of $SO(10)$ at level two, which requires {\it six} Weyl fermions
\dhl (see section (4.1)). Since there is an additional
abelian generator orthogonal to the space spanned
by these roots, the six Weyl fermions actually provide
an embedding of $SO(10)$$\times$$U(1)$.
It is also possible to find
fermionic embeddings of special unitary groups within a
semi-simple group: for example $SU(5)$$\times$$SU(2)$,
with the $SU(5)$ at level two and the $SU(2)$ at level four,
and $Sp(6)$$\times$$SU(3)$, with the $Sp(6)$ at level one
and the $SU(3)$ at level two.

A fermionic realization is a fermionic embedding or representation
together with a realization of the currents and physical states
corresponding to the gauge bosons in a consistent string vacuum.
A fermionic embedding does not necessarily extend
to a fermionic realization,
since we are restricting the constituents of
$CFT_B$ to be real fermions.
A necessary condition is
that one can identify dimension (0,1) operators with
fermionic charges corresponding to all the roots.
For the types of operators in $CFT_A$ which are relevant
for constructing currents, there is a simple relation between
their fermionic charges and their conformal dimension\klt :
\eqn\cdformula{
h = \ha \overrightarrow{Q}{}^2 \qquad .
}
Simple examples are single Neveu-Schwarz fermion operators $\psi$,
$\psi^\dagger$, (which
create single fermionic excitations of the Neveu-Schwarz vacuum)
having $h$$=$$1/2$ and fermionic charge $\pm 1$,
and single Weyl fermion {\it twist} fields $\sigma$, $\mu$,
(which create the doubly-degenerate Ramond vacua from
the Neveu-Schwarz vacuum)
having $h$$=$$1/8$ and fermionic charge $\pm 1/2$.

As will be discussed further in section 5, the $c$$=$$1/2$ conformal
field theory of a single Majorana-Weyl fermion contains
primary fields with conformal dimension
0 (the identity), 1/16 (twist fields), or 1/2 (the Neveu-Schwarz fermion).
Thus there are a limited number of ways to construct currents.
In particular, if $\overrightarrow{Q}$ represents the fermionic charges
corresponding to some root, then the current corresponding to that
root exists only if there is a solution to
\eqn\curdim{
1 = \ha \overrightarrow{Q}{}^2 +
\left( {m_1\over 16} + {m_2\over 2} \right)
}
where $m_1$, $m_2$ are nonnegative integers.

Combining \curdim\ with \levelis , we obtain an important
restriction\myfoot{Condition (i)
of section 5.3 rules out the case $k$$=$$16$.}
on the possible levels for current algebras
with fermionic realizations:
\eqn\kis{
k = 1,~2,~4,~8,~{\rm or}~16 \qquad .
}

It should be noted that the higher level fermionic embedding does
{\it not} uniquely determine the fermionic realization of the
current algebra.
An example is given in the next section.

\newsec{Real Fermionization: examples}

To understand in detail how the constraints from modular
invariance determine the spectrum and couplings of a
solution, it is useful to focus on a specific set of constituent
conformal field theories. Fermionization of the internal
$(2,0)$ unitary conformal field theory is a relatively
straightforward technique for generating explicit solutions
to the string consistency conditions \klt \abk \klst . In this
section we will explain how the ideas we have introduced in the
previous two sections get implemented in the context of
specific examples. These examples have been constructed to
illustrate how particular phenomenological aspects find their
realization in string theory. Although our methodology has the
potential of steadily leading to more phenomenologically
compelling models, the models discussed here were selected for
their pedagogic value only.

The constituent fields of the internal superconformal field theory
are a collection of Majorana-Weyl fermions. Some number of these are
paired into right-moving or left-moving Weyl fermions, or into right-left
paired Majorana (Ising) fermions. The total central charge sums
to $(9,22)$ for a heterotic vacuum with
four dimensional Lorentz invariance.\myfoot{``Heterotic" refers to the
construction of the four dimensional solutions;
it is not necessarily the case that these
solutions possess a large-radius limit which recovers the
ten dimensional heterotic superstring.}
Including the two right-moving Majorana-Weyl fermions with
a spacetime index gives a total of $20$ right-moving and
$44$ left-moving constituent fermions.

The boundary conditions of the fermions about the two non-contractible
loops on the torus specifies their spin-structure. Consider first
the Weyl fermions which are obtained by a complexification
of a pair of Majorana-Weyl fermions, $\lambda(z)=\psi_1(z)+i\psi_2(z)$.
The fermionic charge (bosonic momentum) is allowed to take any
rational value. The possible (twisted) boundary conditions are denoted:
\eqn\complexbcs{\eqalign{
\lambda(\sigma_1 + 2\pi,\sigma_2) &=
-\e{\pi iv}\,\lambda(\sigma_1,\sigma_2) \cr
\lambda^{\dagger}(\sigma_1 + 2\pi,\sigma_2) &=
-\e{-\pi iv}\,\lambda^{\dagger}
(\sigma_1,\sigma_2) \qquad ,}
}
\noindent where $v$ takes any rational value restricted to the domain
$-1$$<$$v$$\le$$1$. The boundary conditions described by eq. \complexbcs\
reduce to a possible sign flip for both
Majorana-Weyl fermions combined
with a rotation of the Majorana-Weyl fermions among themselves:
\eqn\castwor{
\left({\psi_1(\sigma_1 + 2\pi,\sigma_2)\atop
\psi_2(\sigma_1 + 2\pi,\sigma_2)}\right) =
-\left(\matrix{~\,\c(\pi v)&\s (\pi v)\cr
		-\s(\pi v)&\c (\pi v)\cr}\right)
\left({\psi_1(\sigma_1,\sigma_2)\atop
\psi_2(\sigma_1,\sigma_2)}\right) ~.
}
\noindent A right-moving and a left-moving Majorana-Weyl
fermion paired to form a Majorana (Ising) fermion are both
either periodic (Ramond) or antiperiodic (Neveu-Schwarz)
in every sector of the partition function. Any Majorana-Weyl
fermions which are unpaired are called {\it real} fermions.
Real fermions take Ramond or Neveu-Schwarz boundary conditions.

In general, the one-loop vacuum amplitude
(partition function) ${\cal Z}_{\rm Fermion}$
can be written as a sum over all possible spin structures generated
from a set of {\it basis vectors}, $\{ V_i \}$, i.e., the
boundary condition vectors for the constituent fermions
which span the sectors summed over in the partition function:
\eqn\eq{
{\cal Z}_{\rm Fermion}(\tau)=\sum_{\alpha,\beta }
C^{\alpha V}_{\beta V}
{\cal Z}^{\alpha V}_{\beta V}(\tau)\ ,}
where $\{\alpha_i\}$, $\{\beta_i\}$ are
independent sets of nonnegative integers both generating
linear combinations of the basis vectors
vectors $V_i$. The $C^{\alpha V}_{\beta V}$ are projection
coefficients associated with each specification of spin structure;
they determine the phase with which the states in a particular sector
contribute to the partition function.

The ${\cal Z}^{U}_{V}(\tau)$ for
each spin structure are defined in a Hamiltonian representation as:
\eqn\part{
{\cal Z}^{U}_{V}(\tau) =
{\rm Tr} \left\{ (-1)^{{\bf U} \cdot {\widehat {\bf F}}_{\bf V}  }
\;{\rm exp}\left ( 2 \pi i
\tau {\widehat H}_{\bf V}^L - 2 \pi i
\taubar {\widehat H}_{\bf V}^R \right ) \right\} \quad .
}
For the Weyl and Ising  components,
the GSO projection operator, $(-1)^{{\bf U} \cdot {\widehat
{\bf F}}_{\bf V}}$, is defined in the obvious way from
the fermion number operator $\widehat{\bf F}$;
for real fermions its explicit form is more
complicated \klst .

The coefficients $C^{\alpha V}_{\beta V}$
are conveniently rewritten as:
\eqn\Cgeneral{
C^{\alpha V}_{\beta V}=e^{ 2\pi i\left[
-\alpha_i\k\beta_j + \alpha_is_i - \beta_is_i
\right]}\qquad ,
}
where the $k_{ij}$ are rational parameters,
repeated indices are summed,
and $s_i$ takes values
$0$ or $-1/2$, depending on whether the basis vector $V_i$ contributes
spacetime bosons or fermions, respectively.
To define a solution, it is only necessary to specify $k_{00}$ and
the $k_{ij}$ for $i$$>$$j$; the other $k_{ij}$ are then
fixed by modular invariance.

A solution takes the form of a
definite {\it spectrum} of physical states that survive all of
the {\it projections} imposed by string consistency.
The partition functions for interesting solutions sum over
thousands of spin structures, thus it is clearly not practical
to perform the required projections by hand. Instead we
have developed a symbolic manipulation package
\docspec\ which automatically extracts
the massless spectrum of solutions compatible with the
fermionic formulation introduced by Kawai, Lewellen, Schwartz,
and Tye (KLST)\klst . This program takes as input a list of basis vectors,
$V_i$, and projection coefficients, $k_{ij}$.
It then checks for string
consistency, performs the GSO projections, checks for spacetime
supersymmetry, identifies the gauge group and its embedding
from the gauge bosons in the massless spectrum, then outputs
the full massless spectrum organized into irreps of the gauge group.
The tree couplings of physical states can be inferred from their
decomposition into primary fields of the constituent conformal
field theories. However, because of the new formalism required
for real fermions (as will be described in
the next section) we have not yet automated the extraction of
the full tree-level superpotential.

The notion of embeddings makes such a methodology particularly
well-suited to realizing operator algebras that determine
specific spacetime symmetries. Every model contains the
untwisted (i.e. all Neveu-Schwarz) sector,
which ordinarily would contribute the
gauge bosons of the group $SO(44)$, or its regular subgroups.
In the solutions we are interested in, most of the
gauge bosons and chiral matter do {\it not} appear in the
untwisted sector. Rather, the twisted sectors embed most of the
gauge bosons and the matter representations.
This is an important distinction
from the familiar $(2,2)$ compactifications, or $(2,0)$
constructions that are related to $(2,2)$ compactifications  ,
where the low-energy gauge symmetry is realized in the untwisted
sector.

The spin structures are specified by listing the
basis vectors $V_i$, which have 20 right-moving and 44 left-moving
components separated by a double vertical line. Since we use a
64 component Majorana-Weyl notation, Weyl fermion spin structures
are written as left-left or right-right pairs, and Ising fermion
spin structures by left-right pairs.
As always
0,1 denotes Neveu-Schwarz/Ramond boundary conditions; we also use
$++$ and $--$ to denote a Weyl fermion whose boundary condition is
$\mp i$ times itself when taken around a
noncontractible loop.

The first two components of every vector refer to the
right-moving fermions with spacetime indices, $\psi^\mu(\bar{z})$.
Thus (00) in these
slots indicates a spacetime boson;
if $\psi^\mu(\bar{z})$, $X^\mu(\bar{z})$, and $X^\mu(z)$ are not
excited the resulting massless states in such a sector are scalars.
On the other hand, (11) indicates a spacetime fermion,
in this case the two possible values of the ``fermionic
charge'', $\pm1/2$, distinguish the two helicity states.

\subsec{Model A}

This example has $N$$=$$1$ spacetime supersymmetry, $SO(10)$
realized at level two, chiral fermions, and Higgs in the 10 and 45
of $SO(10)$.

\vskip .1in
\halign to \hsbody{\kern-48truept\kern 40pt$#$:\kern .2em&#
&\kern-2.5pt$\Vert$#\cr
V_0&(11111111111111111111&
111111111111$\vert$111111111111111$\vert$111$\vert$11111111111111)
\cr
V_1&(11100100100100100100&
000000000000$\vert$000000000000000$\vert$000$\vert$00000000000000)
\cr
V_2&(00000000000000000000&
111111110000$\vert$111111110000000$\vert$000$\vert$00000000000000)
\cr
V_3&(00000000000000000000&
000000000000$\vert$000011111111000$\vert$000$\vert$00000000000000)
\cr
V_4&(00000000000000000000&
110000111111$\vert$110011001100110$\vert$000$\vert$00000000000000)
\cr
V_5&(11100100010010010010&
111100001100$\vert$101010101010100$\vert$010$\vert$11000000000000)
\cr
V_6&(11010010100100001001&
111100001100$\vert$101001011010011$\vert$101$\vert$00000000000000)
\cr
V_7&(11001001001001100100&
111100001100$\vert$111100001111000$\vert$000$\vert$00110000000000)
\cr
V_8&(00110110110110000000&
000000000000$\vert$010101010101011$\vert$000$\vert$00000000000000)
\cr
V_9&(00000000000000000011&
000000000000$\vert$000000000000000$\vert$011$\vert$001100\p\p\p\p\p\p\p\p)
\cr
}
\centerline{\it Model A}
\goodbreak

\noindent
The $k_{ij}$ for $i$$>$$j$ and $k_{00}$
are all zero except for the following
which are equal to $-1/2$: $k_{71}$, $k_{73}$,
$k_{81}$, $k_{83}$, $k_{85}$, and $k_{86}$.

Apart from the spacetime fermions, the
right-movers in this model
correspond to 7 world-sheet Weyl fermions and 4 Majorana-Weyl
fermions. Three of the Majorana-Weyl fermions ( in slots 17, 19, 20 )
pair up with
left-movers to make 3 Ising fermions; the fourth Majorana-Weyl fermion
( in slot 16 )
is associated with 15 left-moving Majorana-Weyl fermions as a block
of 16 real fermions.
There are 7 fermionic charges associated with the complex right-movers;
they take values 0, $\pm1/2$,
and $\pm 1$ for massless states. These charges result in discrete
symmetries in the low-energy effective theory.

The left movers are separated into four blocks,
embedding the visible matter gauge quantum numbers,
the real fermion spin structures, the Ising fermion spin structures,
and the hidden sector gauge quantum numbers.
In this example the first 12 left-mover slots denote 6 Weyl fermions.
The 6 associated fermionic charges take values 0, $\pm1/2$,
and $\pm 1$ for massless states; these charges
are simply
weights of the visible gauge group $SO(10)$$\times$$U(1)$,
in the basis defined by the embedding of the root lattice
in the sectors which contain the gauge bosons.
The 46 gauge bosons of $SO(10)$$\times$$U(1)$
are distributed in 8 sectors as shown in Table 1.

In the untwisted sector,
massless gauge bosons arise from states with a spacetime fermion
excited and a pair of left-moving Weyl
(or pseudo-complex\myfoot{See section 5.3 for a
discussion of pseudo-complexification.}) fermion
modes excited. In the first 12
left-mover slots which embed $SO(10)$$\times$$U(1)$, there are 66
such pairs, but only six of these survive the projections. These
six gauge bosons correspond to exciting the particle and antiparticle
modes of each of the six Weyl fermions; the resulting fermionic
charges for all six are (0,0,0,0,0,0).
Obviously the six associated
currents are the Cartan elements of $SO(10)$$\times$$U(1)$; because
these Cartan currents are realized by fermion bilinears we can
read off {\it any} weight of $SO(10)$$\times$$U(1)$ from the six
corresponding fermionic charges.

The embedding of $SO(10)$ in these six fermionic charges is
completely characterized by the fermionic charges of
the five simple roots \dhl :

\vskip .2in
\goodbreak
\halign to \hsbody{\kern 60pt(\hfil#,&\hfil#,&\hfil#,&\hfil#,&\hfil#,&
\hfil#)\cr
0&0&0&1&0&0\cr
1/2&-1/2&-1/2&-1/2&0&0\cr
0&0&1&0&0&0\cr
0&1/2&-1/2&0&-1/2&1/2\cr
0&1/2&-1/2&0&1/2&-1/2\cr
}
\vskip .2in
\goodbreak
It is apparent then that the $U(1)$ weight is proportional to the
sum of the fifth and sixth fermionic charges.

There are additional gauge bosons in the untwisted sector which
arise from exciting one of the six Weyl fermions just discussed
together with a mode from one of the seven pseudo-complex left-movers
comprising the block of real fermions. There are 12 distinct
fermionic charges which could result:
($\pm1$,0,0,0,0,0), (0,$\pm1$,0,0,0,0), etc..
However after the GSO projections only four of these appear in
gauge boson states: ($\pm1$,0,0,0,0,0) and (0,$\pm1$,0,0,0,0).

Let us consider the other sectors which contain gauge bosons in turn.
Massless gauge bosons from $V_2$ arise when all the left-movers
are in the vacuum state.
The first 12 left-mover slots of $V_2$ are
(111111110000); the associated fermionic charges are
\eqn\vtwo{
(\pm\half,\pm\half,\pm\half,\pm\half,0,0) \qquad .
}
All of these charges correspond to roots of $SO(10)$,
however, only 8 of these 16 charges appear in gauge boson
states after the projections. The other 8 of these 16 charges
appear in the gauge boson states in $V_2$$+$$V_3$.
Note that $V_2$ and $V_2$$+$$V_3$ differ only by the
boundary conditions of the real fermions, thus it is
the real fermion structure which correlates the GSO projections
in these two sectors.
Massless gauge bosons from $V_3$ require one excited left-moving
Weyl (or pseudo-complex) fermion mode.
The first 12 left-mover slots of $V_3$ are
(000000000000).
There are 12 possible fermionic charges for massless gauge bosons
of $SO(10)$$\times$$U(1)$:
($\pm1$,0,0,0,0,0), (0,$\pm1$,0,0,0,0), etc..
However after the projections only four of these appear in
gauge boson states: (0,0,$\pm1$,0,0,0) and (0,0,0,$\pm1$,0,0).

Massless gauge bosons from $V_4$ arise when all the left-movers
are in the vacuum state;
the associated fermionic charges are
\eqn\vfour{
(\pm\half,0,0,\pm\half,\pm\half,\pm\half) \qquad .
}
Now for a state to be neutral under the extra $U(1)$ of
$SO(10)$$\times$$U(1)$, the sum of the 5th and 6th fermionic
charges must be zero. Thus only 8 of the 16 charges in \vfour\
correspond to roots of $SO(10)$. Of these 8, only four
appear as gauge bosons in $V_4$ after the projections.
The other four appear as gauge boson states in $V_3$$+$$V_4$.
Lastly, the gauge bosons coming from $V_2$$+$$V_4$ and
$V_2$$+$$V_3$$+$$V_4$ are exactly analogous to the above
discussion of $V_4$ and $V_3$$+$$V_4$.
Table 2 summarizes the fermionic charges of the 45
$SO(10)$ gauge bosons.

Thus we have understood the gauge bosons and fermionic charges
corresponding to all 45 roots of $SO(10)$;
this defines an explicit embedding of the gauge group
into 6 fermionic charges.
It is then easy to translate the weights of any other irrep
into fermionic charges, and thus read off the gauge quantum
numbers for all the massless states in the spectrum.
Of course, because of the $N$$=$$1$ spacetime supersymmetry,
the massless matter fields group into chiral supermultiplets
containing a complex scalar and a Weyl spinor. Because the
gravitino resides in sector $V_1$, the superpartner of
a boson/fermion in sector $\alpha_iV_i$ must always be in
sector $V_1$$+$$\alpha_iV_i$.
It is
a convenient shorthand when we count ``states'' in the massless spectrum
to count them four at a time: two scalars and two
CPT conjugate spinor states.

In this model the embedding of $SO(10)$$\times$$U(1)$
is such that fermionic charges
(1/2,1/2,0,0,1/4,-1/4) indicate the highest weight of a 16 of
$SO(10)$, with $U(1)$ charge zero. It is obvious, therefore,
that this model contains
no neutral 16's, since these require boundary conditions
($+$$+$$-$$-$) in left-mover slots 9 through 12. On the other hand,
fermionic charges (1/2,1/2,0,0,1/2,0) indicate the highest weight of a 16 of
$SO(10)$, with $U(1)$ charge 1/2.
Examining the basis vectors we immediately see that sectors
$V_5$, $V_6$, and $V_7$ all potentially contribute states of a massless
16. After performing the projections one finds that in fact
$V_5$ and $V_6$ contribute the highest weights of two chiral 16's
each. However $V_7$ does not contribute any massless states at all
to the spectrum: the projection from $V_9$ removes them. This
feature is independent of the choice of $k_{ij}$'s; it depends
only on the overlap between $V_7$ and $V_9$.

The 16's are chiral because
the helicity is correlated with the $SO(10)$ weight which
distinguishes the 16 from the $\overline{16}$.
One also finds that sector
$V_6+V_8$ contributes the highest weights of two $\overline{16}$'s;
these may couple via adjoint Higgs in sector $V_8$ to the
two 16's in $V_6$, making them superheavy.

It is useful to observe that if the highest weight state of a 16
resides in, say, sector $V_5$, then the states which fill out
this irrep must reside either in $V_5$ or in sectors which are
the sum of $V_5$ and a sector containing $SO(10)$ gauge bosons.
Thus, e.g., for either of the two 16's whose highest weight is
in $V_5$, the full irrep consists of four states from $V_5$
and two states each from $V_2$$+$$V_5$, $V_4$$+$$V_5$,
$V_2$$+$$V_3$$+$$V_5$, $V_2$$+$$V_4$$+$$V_5$, $V_3$$+$$V_4$$+$$V_5$,
and $V_2$$+$$V_3$$+$$V_4$$+$$V_5$. Note that no states of the 16
come from $V_3$$+$$V_5$ in this example, but in general some could.

The full gauge group of this model is
$SO(10)$$\times$$SO(8)$$\times$$[U(1)]^4$.
$SO(8)$ is a hidden sector gauge group and is realized at level one.
However the embedding of $SO(8)$ is nontrivial: the 28 gauge bosons
are distributed in the 16 different sectors which can be formed from
linear combinations of $V_2$, $V_3$, $V_4$, and $2$$*$$V_9$.
Hidden sector massless fields occur in the singlet, $8_v$, $8_s$,
and $8_c$ irreps of $SO(8)$.
The full massless spectrum of
chiral superfields for Model A
is listed in Table 3.
The $U(1)$'s associated with the first two charges listed
are anomalous; the linear combination $2\cdot Q_1$$+$$Q_2$
is anomaly-free.

The role of the block of 16 real fermions in this model is twofold.
First it reduces the rank of the gauge group. The maximal rank for
the gauge group from the left-movers is 22; this is reduced by nine
because of the three Ising fermions and the 15 left-moving
real fermions. Thus the full gauge group has rank 13.

The second role of the real fermions is that they
make it possible to embed
a higher level current algebra, simultaneously producing a
discrete holomorphic algebra. From
the discussion above of the gauge bosons
it is easy to deduce how this model realizes the 45 currents
of $SO(10)$ at level two. The Cartan elements, as already
mentioned, are fermion bilinears
of the form $\lambda^{\dagger}\lambda$
and don't involve
the real fermions.
There are four other currents which are
also fermion bilinears, but where one of the fermions is
pseudo-complex. From
$V_3$ we see that there are four currents which are
composites of one Weyl fermion with 8 real fermion twist fields.
Lastly, there are 32 currents which are composites of
4 Weyl fermion twist fields with 8 real fermion twist fields.

To see the importance of the discrete holomorphic operator algebra,
consider the massless adjoint Higgs in this model. There
are two 45 Higgs supermultiplets in Model A; the scalars
are distributed in sectors as shown in Table 4.

Unlike the gauge bosons, these adjoint Higgs are not associated with
the $SO(10)$ currents, rather they correspond to {\it primary} fields
with respect to the level two $SO(10)$ Kac-Moody current algebra. These
holomorphic primaries
have conformal dimension 4/5. Since the operators which create
physical states must have left conformal dimension 1, the adjoint
Higgs must be a nontrivial
element of the discrete operator
algebra. This is encoded in the real fermion structure of $V_8$.

It is interesting to note that even after fixing the embedding
of $SO(10)$ in fermionic charges, there is still some
residual freedom to adjust the accompanying real fermion
structures. This can be seen by comparing Model A with the
$SO(10)$ level two model of Lewellen \dhl .
Lewellen's model can be obtained from Model A by replacing
$V_5$--$V_9$ with the following:

\vskip .2in
\halign to \hsbody{\kern-48truept\kern 40pt$#$:\kern .2em&#
&\kern-2.5pt$\Vert$#\cr
V_5&(11100100010010010010&
111100001100$\vert$1010101010101010$\vert$1100000000000000)
\cr
V_6&(11010010100100001001&
000000001111$\vert$0000000000000000$\vert$0011110000000000)
\cr
V_7&(00000000101101101101&
111111110000$\vert$0000000000000000$\vert$0000000000000000)
\cr
}
\vskip .2in
\noindent The $k_{ij}$ for $i$$>$$j$ and $k_{00}$
are all zero in this
model.

Lewellen's model embeds $SO(10)$$\times$$U(1)$
into six fermionic charges in
exactly the same way as Model A. However the real fermion
content of the $SO(10)$ currents is slightly different.
In particular, for Lewellen's model the untwisted sector
contributes only the six Cartan gauge bosons, while $V_3$
contributes eight gauge bosons instead of four.
This means that there are
no currents which are
fermion bilinears and where one of the fermions is
pseudo-complex; it also means that
there are eight rather than four currents which are
composites of one Weyl fermion with 8 real fermion twist fields.

Such slight differences in the real fermion structure can
have important consequences for model building.
For example, Model A has a more natural embedding of
$SU(5)$$\in$$SO(10)$ than Lewellen's model. By simply
setting $k_{93}$$=$$-1/2$, the level two $SO(10)$ of
Model A is broken to a level two $SU(5)$, times a $U(1)$.
This is possible because, in Model A, all of the roots of
$SO(10)$ which are {\it not} also roots of $SU(5)$$\times$$U(1)$ are
realized as gauge bosons in sectors involving $V_3$. Modifying
$k_{93}$ causes these gauge bosons to be projected out.
Notice that the central charge of $SU(5)$ at level 2, $c$$=$$48/7$,
is not half-integer valued. Neither is that of the discrete holomorphic
algebra, which has $c$$=$$12-(48/7)$.

\subsec{Model B}

This example has $N$$=$$2$ spacetime supersymmetry, $SO(10)$
realized at level two, and Higgs in the 54 of $SO(10)$.
As in Model A, the five Cartan currents are realized as
simple fermion bilinears in the untwisted sector. However in
Model B these currents are linear combinations of fermion
bilinears corresponding to 10 left-moving Weyl fermions.
The roots of $SO(10)$ are embedded in 10 fermionic charges,
corresponding to the first 20 left-mover slots. The next
16 left-mover slots are again a block of 16 real fermions.

\vskip .2in
\goodbreak
\halign to \hsbody{\kern-48truept\kern 40pt$#$:\kern .2em&#
&\kern-2.5pt$\Vert$#\cr
V_0&(11111111111111111111&
11111111111111111111$\vert$1111$\vert$1111111111111111$\vert$1111)
\cr
V_1&(11100100100100100100&
00000000000000000000$\vert$0000$\vert$0000000000000000$\vert$0000)
\cr
V_2&(00000000000000000000&
11111111000000000000$\vert$0000$\vert$1111111100000000$\vert$0000)
\cr
V_3&(00000000000000000000&
11110000111100000000$\vert$0000$\vert$1111000011110000$\vert$0000)
\cr
V_4&(00000000000000000000&
11110000000011110000$\vert$0000$\vert$1111000000001111$\vert$0000)
\cr
V_5&(00000000000000000000&
11110000000000001111$\vert$0000$\vert$0000000000000000$\vert$0000)
\cr
V_6&(11100100010010010010&
\fm\fm\fm\fm\fm$\vert$\fm$\vert$1100110011001100$\vert$1100)
\cr
V_7&(00000000000000011000&
00000000000000000000$\vert$0000$\vert$0110011001100110$\vert$0110)
\cr
}
\centerline{\it Model B}
\vskip .2in
\goodbreak
\noindent The $k_{ij}$ for $i$$>$$j$ and $k_{00}$
are all zero except for the following
which are equal to $-1/2$: $k_{50}$, $k_{52}$,
$k_{53}$, and $k_{54}$.

The embedding of $SO(10)$ in 10 fermionic charges is
completely characterized by the fermionic charges of
the five simple roots:

\vskip .2in
\goodbreak
\halign to \hsbody{\kern 40pt(\hfil#,&\hfil#,&\hfil#,&\hfil#,&\hfil#,&
\hfil#,&\hfil#,&\hfil#,&\hfil#,&\hfil#)\cr
1/2&1/2&-1/2&-1/2&0&0&0&0&0&0\cr
0&0&1/2&1/2&-1/2&-1/2&0&0&0&0\cr
0&0&0&0&1/2&1/2&-1/2&-1/2&0&0\cr
0&0&0&0&0&0&1/2&1/2&1/2&1/2\cr
0&0&0&0&0&0&1/2&1/2&-1/2&-1/2\cr
}
\ifx\answ\bigans{}\else\vskip .2in\fi
\goodbreak

This model has four Ising fermions; since there are also 16 real
fermions the rank of the full gauge group coming from the
left-moving fermions is 12. There are in addition two $U(1)$
gauge bosons which are part of the $N$$=$$2$ supergravity multiplet;
these states arise in the untwisted sector from exciting a
left-moving spacetime
boson mode and exciting a right-moving Weyl fermion.
Apart from these the full gauge group is
$$SO(10)\times F_4\times SO(5)\times U(1) \qquad ,$$
where the hidden sector gauge group
$F_4$$\times$$SO(5)$ is realized at level one.

The left movers are again separated into four blocks:
the first 20 left-mover slots denote 10 Weyl fermions
whose fermionic charges embed $SO(10)$,
the next 4 are two more Weyl fermions which embed the $U(1)$
and part of $F_4$,
the next 16 are the
real fermions, and the remaining 4 are Ising fermion spin structures.
In this example the embeddings of the visible and hidden
gauge groups overlap: $SO(10)$ is embedded in the first 10
fermionic charges; $F_4$ is embedded in fermionic charges 3 through 8,
11, and 12; and $SO(5)$ is embedded in fermionic charges 1, 2, 9,
and 10.

For Model B the 45 gauge bosons of $SO(10)$
are distributed in 11 sectors as shown in Table 5.

The five Cartan currents are linear
combinations of fermion bilinears of the form
$\lambda^{\dagger}\lambda$.
There are 36 more currents
which are composites of
4 Weyl fermion twist fields with 8 real fermion twist fields.
The remaining 4 currents are composites of 4 Weyl fermion twist fields
with a pseudo-complex fermion from the block of 16 real fermions.
These 4 currents correspond to the gauge boson states which arise
in $V_5$.
Table 6 summarizes the fermionic charges of the 45
$SO(10)$ gauge bosons.

Because of the $N$$=$$2$ spacetime supersymmetry,
the massless spectrum assembles into $N$$=$$2$
supermultiplets. Apart from the supergravity multiplet,
there are 2608 massless states which belong
to supermultiplets containing either (i) a gauge boson, two Weyl
spinors, and a complex scalar, or (ii) two Weyl spinors and
two complex scalars. Thus the supermultiplets
containing the 108 gauge bosons of
$SO(10)$$\times$$F_4$$\times$$SO(5)$$\times$$U(1)$
account for 864 states; the remaining states form 218
matter supermultiplets
in the following irreps:

--- one 54 of $SO(10)$,

--- one 26 of $F_4$,

--- one 5 of $SO(5)$,

--- four pairs $16$$+$$\overline{16}$ of $SO(10)$
which also carry charge $1/4$,
$-1/4$ respectively under the $U(1)$,

--- a pair which carry only $U(1)$ charge $\pm1$, and three
which are singlets under the full gauge group.

For $SO(10)$ at level two, the 54 and the 45 are the only new
irreps which can occur as massless matter states other than the
irreps which also occur at level one (the singlet, 10, 16,
and $\overline{16}$). As was discussed above, a 45 Higgs corresponds
to a level two Kac-Moody primary with conformal dimension
4/5, and must therefore be a nontrivial element of the
discrete algebra.
A 54 Higgs corresponds to a level two Kac-Moody
primary with conformal dimension 1; since the full physical
vertex operator also has left conformal dimension 1, this implies
that it must be the identity element under the discrete algebra.
It is not surprising then that the states of the 54 arise in precisely
the same sectors as the $SO(10)$ gauge bosons, which are also trivial
under the discrete algebra.
Moreover, if we construct Table 7 listing the sectors and
fermionic charges of the (scalar) states in the 54, it differs
from Table 6 {\it only} by the states in the untwisted sector.

The highest weight states of the (nonchiral) 16's arise in
sectors $3$$*$$6$ or $3$$*$$6$$+$$7$, reflecting that fact
that with this embedding of $SO(10)$ the highest weight of
the 16 is given by
$$ ({1\over 4},{1\over 4},{1\over 4},{1\over 4},{1\over 4},
{1\over 4},{1\over 4},{1\over 4},{1\over 4},{1\over 4}) \qquad .$$

There are many variations of Model B which preserve the
realization of $SO(10)$ at level two. For example, we can
add the following additional basis vector:

\vskip .2in
\goodbreak
\halign to \hsbody{\kern-48truept\kern 40pt$#$:\kern .2em&#
&\kern-2.5pt$\Vert$#\cr
V_8&(11001001001001100100&
00000000000011110000$\vert$1111$\vert$0000000000000000$\vert$0000)
\cr
}
\vskip .2in
\goodbreak
\noindent The additional $k_{ij}$ for $i$$>$$j$ are chosen
to be all zero except
for $k_{84}$$=$$-1/2$, and $k_{86}$$=$$1/4$.

For this model the $N$$=$$2$ spacetime supersymmetry of Model B is
broken to $N$$=$$1$. The full gauge group is given by
$$SO(10)\times Sp(6)\times SO(5)\times SU(2)\times U(1) \qquad ,$$
\noindent which is again rank 12. The $SO(10)$ is realized at
level two, and the other factors at level one.

In closing this section on examples we should emphasize that
our symbolic manipulation package makes the construction and
analysis of such models quite easy. All of the results presented
here come directly from the computer printout, and were produced
in approximately one minute on a {\sl NeXT}.
Anyone who has gained some
familiarity with the modular invariance constraints could
produce and analyze dozens of variations on Models A and B in a single
afternoon.

\newsec{Aspects of real fermionization}

\subsec{Tree-level Couplings}

The tree-level correlation functions of the $N$$=$$(2,0)$ superconformal
field theory are an essential ingredient in extracting the full
tree-level superpotential of the low-energy effective field theory.
Any solution to string theory that realizes a higher level current
algebra must, if it
has a fermionic embedding, {\it necessarily} contain some number of
real fermion constituents, i.e., Majorana-Weyl fermions which
cannot be paired into either Ising or Weyl fermions
in every sector of the partition function. The correlators of a real
fermion conformal field theory cannot be abstracted from those of the
critical Ising model or of free bosons, and thus require an independent
analysis.

In the fermionic construction given by Kawai, Lewellen, Schwartz, and Tye
(KLST), any three sectors of the partition function allow a
{\it pseudo-complexification}: a pairing of the real fermions
that is consistent with their boundary conditions in each of the three
sectors \klst . This property of their construction is motivated by requiring
modular invariance of non-vanishing two loop amplitudes in the
factorization limit.
Conservation of the
pseudo-U(1) charges associated with such pseudo-complexifications
then provides a {\it prescription} for computing arbitrary
3-point and 4-point correlators involving real fermions. However
even this prescription breaks down for general $N$-point
correlators, $N$$>$$4$.
Clearly, it would be useful to have a more complete understanding of
real fermion conformal field theories, both as a consistency
check on the limits of the validity of the KLST prescription, and with a view
towards developing direct tree-level methods that can be extended to
other cases of interest.

Let us consider an alternative starting point. For rational conformal
field theories, such as real fermions, Verlinde's theorem \ver\ allows us
to make explicit contact between the modular transformation properties
of the chiral spin structure blocks in the one-loop partition function, and
the tree-level fusion algebra of the chiral primary field operators.
The correspondence works as follows. In a rational conformal field
theory it is possible to rewrite the one-loop partition
function in terms of a finite number of holomorphic
blocks, $\chi _i(\tau)$, which are the characters of the
chiral primary fields, $\phi_i(z)$, under the Virasoro algebra
(or an extension thereof).
Using the characters, one can form a suitable basis for the
action of the modular transformations, $S: \tau$$\to$$-1/\tau$, and
$T : \tau$$\to$$\tau$$+$$1$, such that
$S$ and $T$ are realized as finite dimensional unitary matrices.
It is easy to show that if the characters are modular functions the
matrices $S$ and $T$ satisfy two important consistency conditions:
\eqn\stcon{
(ST)^3 = S^2 = C\qquad .
}
Here $C$ is the {\it conjugation} matrix that takes each
character to its conjugate, and satisfies $C^2$$=$${\bf 1}$, the
unit matrix. The existence of a conjugation matrix is related to the fact
that in the tree-level operator product algebra, every chiral primary
field operator is associated with a unique conjugate:
let $[\phi_i]$, $[\phi_i^c]$ denote the conformal families whose chiral
primary fields are $\phi_i$ and $\phi_i^c$, respectively, and let
$[\i]$ denote the conformal family of the identity operator. Then
\eqn\conj{
[\phi_i ] \times [ \phi_i^{c}] ~~=~~ {[\i]} ~~~~~~,
}
\noindent defines the chiral primary field operator, $\phi_i^c$,
conjugate to $\phi_i$. Of course an operator could be
{\it self}-conjugate.
Verlinde's theorem is the statement that the matrix $S$,
derived in an appropriate basis from the characters,
diagonalizes (and determines) the tree-level fusion rules.
Let the subscript `$0$' denote the conformal family of the
identity operator, $\i$. Note that in a unitary conformal field theory
the identity is the unique operator with conformal dimension zero.
Construct
\eqn\fuscoef{
\N = \sum_n {S_{in}S_{jn}S_{nk}\over S_{0n}} ~~~~~~,
}
where the coefficients $\N$ are nonnegative integers.
The fusion rules are then given by
\eqn\fusion{
[\phi_i] \times [\phi_j] = N_{ijl}C^{lk}[\phi _k] ~~~~~.
}
\noindent The $\N$ also give selection rules on the
3-point chiral correlators since
\eqn\threepoint{
\vev{\phi_i(z_1)\phi_j(z_2)\phi_k(z_3)} \propto \N ~~~~~.
}

A single left-moving Majorana-Weyl fermion corresponds
to a $c_L$$=$$1/2$ conformal field theory. The Virasoro primaries
have conformal dimension 0 (the identity, $\i$), 1/2 (the chiral
fermion field $\psi(z)$), or 1/16 (the chiral twist fields).
In general (see \ginsparg )
there may be two distinct chiral
twist fields $\sigma(z)$ and $\mu(z)$; this is the case if we require
the existence of a well-defined chiral fermion number, i.e. an
operator $(-1)^{F_L}$ which anticommutes with $\psi(z)$:
\eqn\antic{
\left\{ (-1)^{F_L},\psi_n \right\} = 0
}
for all modes $\psi_n$. Acting on the Neveu-Schwarz vacuum $\ket{0}$,
$\sigma(0)$ and $\mu(0)$ create two degenerate Ramond vacua with
different fermion number. The Ramond zero mode operator $\psi_0$,
$(\psi_0)^2$$=$$1/2$,
takes one Ramond vacuum into the other. This implies the obvious
fusion rule
\eqn\obfr{
[\psi]\times[\sigma] = [\mu]\qquad .
}

To apply Verlinde's theorem, the chiral spin
structure blocks of the one-loop partition function should be
rewritten in terms of the {\it four} chiral Virasoro characters
$\chi_0$, $\chi_{\sigma}$, $\chi_{1/2}$, and $\chi_{\mu}$.
Of course the Virasoro characters $\chi_{\sigma}(\tau)$ and
$\chi_{\mu}(\tau)$ are actually equal, since the corresponding
primaries have the same left conformal dimension.
We write \ginsparg
\eqn\ourchar{\eqalign{
{\cal Z}_0^0(\tau) ~&\equiv ~ \chi_0(\tau) ~+~ \chi_{1/2}(\tau) \cr
{\cal Z}_{1}^0(\tau) ~&\equiv ~ \chi_0(\tau) ~-~ \chi_{1/2}(\tau) \cr
{\cal Z}_0^{1}(\tau)
  &\equiv ~ \chit_{\sigma}(\tau) ~+~ \chit_{\mu}(\tau)\cr
{\cal Z}_{1}^{1}(\tau)
  &\equiv ~ \chit_{\sigma}(\tau) ~-~ \chit_{\mu}(\tau) \quad ,
}}
where we have introduced the notation
$\chit_{\sigma}$$\equiv$$\chi_{\sigma}$$/$$\sqrt{2}$,
$\chit_{\mu}$$\equiv$$\chi_{\mu}$$/$$\sqrt{2}$.
If we use the basis
$\chi_0$, $\chi_{\sigma}$, $\chi_{1/2}$, $\chi_{\mu}$,
to construct $S$, then $S$ will not be unitary; this
reflects the fact that one does not obtain a {\it diagonal}
modular invariant using {\it all four} characters.
We have adapted Verlinde's analysis to this case, however here we
will employ the convenient shortcut of using the modified
basis $\chi_0$, $\chit_{\sigma}$, $\chi_{1/2}$, $\chit_{\mu}$.

Since the Ramond-Ramond block
${\cal Z}_{1}^{1}(\tau)$ vanishes, it may not seem that its
modular transformation properties under $S$ and $T$ are
meaningful. However it is apparent in the KLST formalism
that ${\cal Z}_{1}^{1}(\tau)$ picks up phases under $S$ and $T$,
and that these phases are vital to the construction of the
partition function for real fermions. In
\klst\ this was understood by appealing to higher loop
modular invariance: although ${\cal Z}_{1}^{1}(\tau)$ vanishes,
it appears in the factorization limit of certain {\it nonvanishing}
two-loop amplitudes. Here we see that the modular transformation
properties of ${\cal Z}_{1}^{1}(\tau)$ are needed to connect the
one-loop partition function to the tree-level fusion rules.
Both arguments may be regarded as appealing to the
unitarity of the internal rational conformal field theory.
To be completely general, we will parametrize the modular
transformations of ${\cal Z}_{1}^{1}(\tau)$ by two phases:
\eqn\ourmod{\eqalign{
\tau\to -1/\tau :\quad
&\;\;{\cal Z}^{0}_{0} \to {\cal Z}^{0}_{0}\;\cr
&\;\;{\cal Z}^{0}_{1} \to {\cal Z}^{1}_{0}\;\cr
&\;\;{\cal Z}^{1}_{0} \to {\cal Z}^{0}_{1}\;\cr
&\;\;{\cal Z}^{1}_{1} \to \e{i\phi}{\cal Z}^{1}_{1}\cr
\tau\to\tau +1 :\quad
&\;\;{\cal Z}^{0}_{0} \to \e{-{\pi i\over 24}}{\cal Z}^{0}_{1}\;\cr
&\;\;{\cal Z}^{0}_{1} \to \e{-{\pi i\over 24}}{\cal Z}^{0}_{0}\;\cr
&\;\;{\cal Z}^{1}_{0} \to \e{{\pi i\over 12}}{\cal Z}^{1}_{0}\;\cr
&\;\;{\cal Z}^{1}_{1} \to
\e{{i\eta}}\e{{\pi i\over 12}}{\cal Z}^{1}_{1} \qquad .
}}

The parameters $\phi$ and $\eta$ are then fixed by combining
\ourchar\ with \ourmod\ and imposing the consistency conditions
\stcon . Thus requiring $(ST)^3$$=$$S^2$ gives
\eqn\etais{
\eta = {\pi\over 12}-{\phi\over 3} \qquad .
}
\noindent The constraint $S^4 $$=$${\bf 1}$ has {\it two distinct
solutions}:
$$\phi = 0,\quad{\pi\over 2} \qquad . $$

We thus obtain two possible forms for $S$ acting as a $4$$\times$$4$
unitary matrix on the modified basis set
$\chi_0$, $\chit_{\sigma}$, $\chi_{1/2}$, and $\chit_{\mu}$:
\eqn\rtypes{\eqalign{
\phi = 0:\quad &S =
\ha\left(\mymatrix{1&{\hfil 1}&{\hfil 1}&{\hfil 1}\cr
		 1&{\hfil 1}&{\hfil -1}&{\hfil -1}\cr
		 1&{\hfil -1}&{\hfil 1}&{\hfil -1}\cr
	 	 1&{\hfil -1}&{\hfil -1}&{\hfil 1}\cr}\right) \cr
\phi = {\pi\over 2}:\quad &S =
\ha\left(\mymatrix{1&{\hfil 1}&{\hfil 1}&{\hfil 1}\cr
		 1&{\hfil i}&{\hfil -1}&{\hfil -i}\cr
		 1&{\hfil -1}&{\hfil 1}&{\hfil -1}\cr
	 	 1&{\hfil -i}&{\hfil -1}&{\hfil i}\cr}\right) \qquad .
}}
Verlinde's theorem then provides the corresponding tree-level
fusion rules:
\eqn\ourfu{\eqalign{
\phi = 0:\quad
[\psi] \times [\psi] &= [\i] \cr
[\psi] \times [\sigma] &= [\mu] \cr
[\sigma] \times [\sigma] &= [\i]\cr
[\mu] \times [\mu] &= [\i]\cr
[\sigma] \times [\mu] &= [\psi]\cr
\phi = {\pi\over 2}:\quad
[\psi] \times [\psi] &= [\i] \cr
[\psi] \times [\sigma] &= [\mu] \cr
[\sigma] \times [\sigma] &= [\psi]\cr
[\mu] \times [\mu] &= [\psi]\cr
[\sigma] \times [\mu] &= [\i] \qquad .
}}
We will refer to the $\phi$$=$$0$ case as the {\it s-type} fusion rules,
for self-conjugate twist fields, and the  $\phi$$=$$\pi/2$ case as the
{\it c-type} fusion rules. In the latter fusion algebra the twist fields
are conjugates of each other.

Our result is that in any solution obtained via real fermionization each
constituent real fermion can be labelled as s-type or c-type, where
this {\it labeling} denotes the corresponding set of fusion rules. It is
important to
realize that this should {\it not} be regarded as a new result in the
conformal field theory of free Majorana-Weyl fermions {\it per se},
rather it is a new result about the proper conformal field theory
interpretation of solutions to string theory obtained in the
fermionic formulation.

To emphasize this last point, we sketch how to recover the
familiar fusion rules of the Ising model.
The critical Ising model does not require the
existence of a chiral $(-1)^{F_L}$, only of the non-chiral combination
$(-1)^F$$=$$(-1)^{F_L+F_R}$. Thus for the Ising model we need introduce
only a single chiral twist field $\sigma^+(z)$, where
$\sigma^{\pm}(z)$$=$$(\sigma(z)$$\pm$$\mu(z))$$/$$\sqrt{2}$.
The unitary matrix $S$ is now computed in the new basis
provided by the four chiral Virasoro characters
$\chi_0$, $\chi_{\sigma^+}$, $\chi_{1/2}$,
and $\chi_{\sigma^-}$. The result is identical for the s-type
and c-type cases:
\eqn\isingS{
S = \ha\left(\matrix{1&\sqrt{2}&1&0\cr
		\sqrt{2}&0&-\sqrt{2}&0\cr
	 	  1&-\sqrt{2}&1&0\cr
		  0&0&0&2\cr}\right) \qquad .
}

Clearly $\sigma^-(z)$ decouples; it can be consistently set to zero.
Application of Verlinde's theorem then gives the fusion rules:
\eqn\isingfu{\eqalign{
[\psi] \times [\psi] &= [\i] \cr
[\psi] \times [\sigma] &= [\sigma] \cr
[\sigma] \times [\sigma] &= [\i] + [\psi] \qquad ,
}}
where the superscript $+$ on $\sigma$ has been dropped. These are
the familiar fusion rules appearing in, e.g., \ver .

\subsec{Selection Rules}

Given explicit fusion rules for the chiral
primaries of the real fermions the correlators can be obtained
via the conformal bootstrap.
We intend to give a complete treatment of such
computations in future work.
A useful means of finding selection rules for correlators
is to introduce the notion of {\it simple currents} (also called bonus
currents), discussed for general rational conformal field
theories in \ki \sy .
A simple current is defined as any chiral primary
$\phi_i(z)$ in the
chiral operator product algebra such that
\eqn\simpdef{
\sum_k\,N_{ij}^k = 1\;,\quad{\rm for~all~}j.
}
For example, in the Ising fusion rules \isingfu , $\psi(z)$ is
a simple current, but $\sigma(z)$ is not.

In general simple currents are not currents, i.e. they need not
have conformal dimension $=$$1$. However associated with each
simple current is a discrete symmetry, and a corresponding charge
which is conserved $mod~1$ in correlators. This is easy
to demonstrate for the fusion algebras \ourfu\ obtained above. For any
simple current $\phi_i(z)$, there must be a positive integer $N$
such that $[(\phi_i)^N]$$=$$\i$. $N$ is called the {\it order} of
the simple current. Thus for example in the s-type algebra
\ourfu , $\sigma(z)$ is a simple current of order 2, while in the
c-type algebra $\sigma(z)$ is a simple current of order 4.
Clearly the chiral primaries of any rational fusion algebra can be
decomposed into orbits with respect to each simple current.
Thus in the s-type algebra, the orbits with respect to $\sigma(z)$
are $\{\i$,$\sigma\}$, $\{\mu$,$\psi\}$; for the c-type algebra,
there is only one orbit: $\{\i$,$\sigma$,$\psi$,$\mu\}$.

For any simple current $\phi_i(z)$, there is a discrete charge $Q_j$
assigned to every primary $\phi_j(z)$. When the matrix $S$ is
symmetric (as in \rtypes ), these charges are given by the simple
expression\sy :
\eqn\chargeis{
\e{2\pi i Q_j} = {S_{ij}\over S_{oj}}\qquad .
}

These charges are conserved $mod~1$ in correlators. This
provides useful selection rules for $N$-point functions
involving real fermions. One of these selection rules is already
familiar: $\psi(z)$ is a simple current with an associated
$Z_2$ charge. This charge is the same for the s and c-type algebras.
Conservation of this charge gives the selection rule that correlators
with an {\it odd} number of Ramond fields vanish \joel .

\subsec{Consistency of the KLST Construction}

The analysis of the previous section makes an explicit connection
between the one-loop partition function of real fermions, and the
tree-level operator algebra of the underlying conformal
field theory. This allows us to perform some consistency
checks on the KLST formulation \klst . We will show that
for a large class of consistent solutions,
the prescription given in \klst\ is both
necessary and sufficient. However we will also derive the
simplest case where the KLST prescription
appears to break down.
The problem can be traced
to the assumed modular tranformations of the real fermion
spin structure blocks.

The KLST prescription includes three constraints
which apply only to the real
fermion spin structures in the partition function.
These are \klst :

\noindent(i) The total number of real fermions is even.

\noindent(ii) Let $O(V_i,V_j)$ denote the number of overlaps
of real fermions with the Ramond boundary condition between sectors $V_i$
and $V_j$. Then for all $V_i$, $V_j$, $O(V_i,V_j)$ must be even.

\noindent(iii) Let $O(V_i,V_j,V_k)$ be the number of overlaps of
real fermions with the Ramond boundary condition common to three sectors.
Then for all $V_i$, $V_j$, $V_k$, $O(V_i,V_j,V_k)$ must be even.
This is referred to as the cubic constraint in \klst \joel \dhl .
Note that, since the all-Ramond basis vector $V_0$ is in
every model, (ii) is actually implied by (iii). By the same
token, $O(V_0,V_i)$ even implies that the total number of real
fermions with the Ramond boundary condition in any single
basis vector must be even.

The KLST construction relies on pseudo-complexification
of pairs of real fermions in order to define the Fock space
upon which the GSO projection operators act. Pseudo-complexification
means that, in every sector, real fermions are sorted ---in a
sector-dependent way---
into NS-NS or R-R pairs. Each pair is then used to
define a complex fermion, and the Fock space is constructed
as if these complex fermions were actual Weyl fermions.
The resulting Fock space is obviously a subspace of the
original Fock space spanned by the real fermions.

The KLST construction also relies on the pseudo-complexification of
pairs of real fermions in order to define the modular transformation
properties of the chiral spin structure blocks of a single real
fermion. The transformation properties were assumed to be given
(up to a sign) by
the ``square root'' of those for a Weyl fermion. Thus
\eqn\modreal{\eqalign{
\tau\to -1/\tau :\quad
&{\cal Z}^{0}_{0} \to {\cal Z}^{0}_{0}\;\quad
{\cal Z}^{0}_{1} \to {\cal Z}^{1}_{0}\;
\cr
&{\cal Z}^{1}_{0} \to {\cal Z}^{0}_{1}\;\quad
{\cal Z}^{1}_{1} \to \e{{\pi i\over 4}}{\cal Z}^{1}_{1}
\cr
\tau\to\tau +1 : \quad
&{\cal Z}^{0}_{0} \to \e{-{\pi i\over 24}}{\cal Z}^{0}_{1}\;\quad
{\cal Z}^{0}_{1} \to \e{-{\pi i\over 24}}{\cal Z}^{0}_{0}\;
\cr
&{\cal Z}^{1}_{0} \to ~ \e{{\pi i\over 12}}{\cal Z}^{1}_{0}\;\quad
{}~ {\cal Z}^{1}_{1} \to  ~ \e{{\pi i\over 12}}{\cal Z}^{1}_{1}
\qquad .}
}

One immediately notes that this does not agree with
the modular transformation properties of either the s-type
or the c-type cases discussed above. However in a partition
function of $N$ real fermions, the modular transformations
of relevance are those of the real fermion spin structure blocks
{\it taken $N$ at a time}. Suppose that in a particular sector of the
partition function, there are $N_s$, $N_c$ left-moving
real fermions with Ramond boundary condition and fusion algebra
of s, c type, and $\br$, $\bi$ right-moving real fermions with
Ramond boundary condition and fusion algebra of s, c
type. According to the transformation properties under $S$
assumed in the KLST prescription \modreal ,
the corresponding real fermion spin structure blocks transform by
the overall phase
\eqn\klstph{
{\rm exp}{\pi i (N_s + N_c - \br -\bi)\over 4} ~~~~~~.
}
Our analysis in the previous section indicates that the overall
phase should be
\eqn\ourph{
{\rm exp}{\pi i (N_c  -\bi)\over 2} ~~~~~~~.
}
Thus consistency between the two prescriptions for the
modular transformation properties is achieved if and only if
\eqn\need{
(N_s+\bi ) - (N_c+\br ) = 0 \quad {\rm mod~8}.
}
{\it for every sector} in the partition function. Since the chiral
spin structure blocks of right-moving c-type real fermions transform
like those of left-moving s-type real fermions for the purposes of
this argument, we will suppress the left-right labeling and write simply
\eqn\needs{
N_s - N_c = 0 \quad {\rm mod~8.}
}
This is the basic identity required for agreement
between the assumed modular transformation properties in the KLST
prescription, and those derived from the tree-level
fusion rules of the real fermion conformal field theory.

Our task now is to convert this consistency equation into a list of constraints
on the {\it basis vectors}. i.e., the set of boundary condition vectors
which span the sectors of the partition function. One obvious consequence
of \needs , given that the sector $V_0$ occurs in any solution,
is that the total number of real fermions in the underlying
conformal field theory must be even (thus reproducing (i) above).
In a sector where $N_s$$=$$N_c$ ({\it not} merely mod 8),
there are as many real fermions with Ramond boundary condition
and fusion algebra of s-type as of c-type, and as many real fermions with
Neveu-Schwarz boundary condition and fusion algebra of s-type as of
c-type. Thus we have a collection of s-c pairs. However a Weyl fermion with
periodic or antiperiodic boundary condition may also be regarded as an s-c
pair of real fermions: the holomorphic operator algebra of a Weyl
fermion is a subalgebra of that obtained from the tensor product
of an s-type algebra and a c-type algebra, with only 4 chiral
primaries instead of the possible $4$$\times$$4 = 16$. Thus in any
sector where $N_s$$=$$N_c$ we can perform a sector dependent
pseudo-complexification of the real fermions. This is the essence
of the KLST prescription for real fermions.

Let us now suppose that the constraint \needs\ is satisfied
by the set of basis vectors and derive what additional
constraints may follow by requiring \needs\ for sectors which
are sums of basis vectors. To do this, let $R(V_1$$+$$V_2$$+$$...$$+$$V_k)$
denote the number of real Ramonds in the sector defined by the
sum of basis vectors $V_1$$+$$V_2$$+$$...$$+$$V_k$. Then one
can easily verify the following identity:
\eqn\iden{\eqalign{
R(V_1+V_2+...+V_k&) = \sum_{i}\,R(V_i) - 2\sum_{i<j}\,
O(V_i,V_j)\cr
&+4\sum_{i<j<k}\, O(V_i,V_j,V_k)
-8\sum_{i<j<k<l}\, O(V_i,V_j,V_k,V_l)+\ldots
}}

Applying \needs\ and \iden\ to the sum of any two basis vectors,
one finds:
\eqn\ourtwo{
O_s(V_i,V_j) - O_c(V_i,V_j) = 0 \quad{\rm mod~}4,
}
where $O_s$ and $O_c$ denote the numbers of overlaps of real fermions with
Ramond boundary condition and s-type or c-type fusion algebra, respectively.
Since $O(V_i,V_j)=O_s(V_i,V_j)$$+$$O_c(V_i,V_j)$, \ourtwo\ implies
constraint (ii). However \ourtwo\ is a somewhat stronger
constraint than (ii).

Applying \needs\ and \iden\ to the sum of any three basis vectors,
one finds:
\eqn\ourcub{
O_s(V_i,V_j,V_k) - O_c(V_i,V_j,V_k) = 0 \quad{\rm mod~}2.
}
This is obviously equivalent to the cubic constraint (iii).

Applying \needs\ and \iden\ to the sum of any four basis vectors,
one finds:
\eqn\ourquart{
O_s(V_i,V_j,V_k,V_l) - O_c(V_i,V_j,V_k,V_l) = 0 \quad{\rm mod~}1.
}
However this is no constraint at all, since $O_s$ and $O_c$ are
integers. There is therefore no ``quartic constraint''
for real fermions, a fact
which was first obtained by KLST \klst .
Similarly looking at sums of $>4$ basis vectors
produces no additional constraints.

\subsec{Spin Structures For Real Fermions}

So far we have shown that the consistency condition
\needs\ suffices to derive the KLST constraints (i)-(iii) {\it without}
making any reference to higher-loop modular invariance. To see whether
\needs\ implies any {\it additional} requirements beyond (i)-(iii), we
will consider the general form of sets of basis vectors which
describe real fermion spin-structures. We will suppress the entries of a basis
vector which describe Weyl or Ising fermions, writing
$N$ dimensional basis vectors, where $N$ is the number of real
fermions. We can also suppress the distinction between left-movers
and right-movers for the purposes of this argument. The real
fermions are of course either periodic or antiperiodic. Furthermore,
the boundary conditions have been chosen such that there are no
{\it global pairs}, i.e. no two real fermions have identically matched
boundary conditions across the entire set of basis vectors. Obviously
such a pair should have been regarded as a single Weyl or Ising fermion
and thus (by assumption) suppressed.

We have already shown that the KLST constraints (i)-(iii) will
follow provided that \needs\ is satisfied for any {\it basis vector},
and that \ourtwo\ is satisfied for any two basis vectors.
Thus our strategy will be to construct sets of basis vectors
which describe real fermions and also satisfy constraints (i)-(iii).
The set of basis vectors therefore defines a solution to string theory
built consistent with the KLST prescription. We then need to show that
for any such set of basis vectors, there
exists at least one s-c labeling of the N real fermions such that
\needs\ and \ourtwo\ are satisfied. It follows that there is an
unambiguous definition of the tree-level fusion rules for all of
the real fermions. In each case where at least one s-c labeling exists,
the KLST constraints (i)-(iii) are not only necessary but also sufficient.

Consider a set of $M$ basis vectors describing the spin structure of
$N$ real fermions. We will consider these as $N$ dimensional vectors
whose entries are either 0 (denoting Neveu-Schwarz) or 1 (denoting Ramond).
For simplicity we may always assume that we have a {\it minimal set} of
basis vectors, in the sense that if any one basis vector were to be
removed, at least two real fermions would become globally paired.
We will not bother to write $V_0$, the basis vector with all real
fermions in the Ramond ground state, which is always present.
Applying constraints (i)-(iii), we then derive the following results:

1. For $M$$\leq$$3$, there are no allowed sets of basis vectors which
contain real fermions.

2. For $M$$=$$4$, there is a {\it unique} set of basis vectors
(modulo relabeling or reshuffling the basis) which contains
real fermions. This unique set of four produces $16$ real
fermions:

\goodbreak
\halign to \hsbody{\kern 20pt$#$: &#\cr
V_1&(1111111100000000)
\cr
V_2&(1111000011110000)
\cr
V_3&(1100110011001100)
\cr
V_4&(1010101010101010)
\cr
}
\goodbreak

\noindent The proof is as follows. In a collection of four vectors
as above, each {\it vertical}
column is a 4-digit binary number from 0000 to 1111.
To avoid any global pairing, any particular 4-digit binary must
appear just once or not at all. Thus the {\it maximum} number of
real fermions which we can describe with four basis vectors is
clearly 16. Now consider the column 1111 (the first column above).
It is easy to see that if 1111 is present, then constraints (i)-(iii)
imply that all 16 columns must be present. On the other hand, if
1111 is absent, then (i)-(iii) have no solutions. Thus 16 is also the
minimum number of real fermions, and this is in fact the unique
allowed spin structure.

3. There are many s-c labelings of the structure of 16 which satisfy
\needs\ and \ourtwo . Two examples are
\eqn\labeling{
\eqalign{
&scscscscscscscsc\cr
&ssssccccsssscccc \qquad .
}}

4. It is not difficult to show \feng\ that 16 is the minimum number
of real fermions {\it for any} $M$.

5. For $M$$=$$5$, the allowed spin structures
describe either 16 or 32 real fermions. For a collection of
five basis vectors, each vertical column is a 5-digit binary
between 00000 and 11111. Thus 32 is the maximum number of real
fermions which can be produced, and in fact this unique structure
of 32 also satisfies the constraints (i)-(iii). It can be
written as

\goodbreak
\halign to \hsbody{\kern 20pt$#$: &#\cr
V_1&(11111111000000001111111100000000)
\cr
V_2&(11110000111100001111000011110000)
\cr
V_3&(11001100110011001100110011001100)
\cr
V_4&(10101010101010101010101010101010)
\cr
V_5&(11111111111111110000000000000000)
\cr
}
\goodbreak

\noindent This form makes it clear that the structure of 32 consists
of two copies of the structure of 16. The fifth basis vector merely
breaks the symmetry between the two blocks of 16. Thus to get an
allowed s-c labeling for the structure of 32, we merely take any two
of the allowed labelings for the structure of 16.

To complete the discussion of $M$$=$$5$, we
note that the constraints
(i)-(iii) are all $mod~2$ constraints. It follows
immediately that if there is any spin structure satisfying (i)-(iii)
and describing $N$ real fermions, then there exists another allowed
structure which describes $32$$-$$N$ real fermions. This second ---or
``complement''---
structure is obtained from the first by simply removing the columns
which appear in the first structure from the structure of 32 above.
Thus there are also no allowed structures with $16$$<$$N$$<32$.

6. For $M$$>$$5$, the classification of allowed spin structures
for real fermions gets more complicated. For example, for $M$$=$$6$,
an exhaustive search shows that there are allowed structures for
16, 24, 28, 32, 36, 40, 48, and 64 real fermions. The structure of 64
is maximal, and may be regarded as four blocks of 16. The structures
with 36, 40, and 48 real fermions are $64$$-$$N$ complements of
the structures which give 28, 24, or 16 real fermions. Thus the only
essentially new
structures are those
giving 24\myfoot{This structure of 24 was derived and pointed
out to us by Jonathan Feng, who has also found a different
structure of 28 for $M$$=$$7$.}
or 28 real fermions.
The structure of 24 may be thought of as two
{\it overlapping} blocks of 16, and inherits a number of allowed
s-c labelings from those of the 16.
More generally, although we have not completed the classification
of all allowed spin structures for $M$$>$$5$, it is clear that
a large class of the allowed structures are built from the basic block
of 16, and furthermore that they inherit
allowed s-c labelings in an obvious way from the component blocks.

7. The structure of 28 real fermions
for $M$$=$$6$ is more interesting. It can
be written as
\goodbreak
\vbox{\offinterlineskip
\halign to \hsbody{\kern 20pt\strut#&#&#&#&#\cr
\omit\kern 20pt&\kern 4pt\ru&\ru&&\cr
$V_1$: &(\vr11111111&\kern 0.4pt00000000\vr&00000010&1110)
\cr
$V_2$: &(\vr11110000&\kern 0.4pt11110000\vr&00000011&1001)
\cr
\omit\kern 20pt&&\ru&\ru&\cr
$V_3$: &(\vr11001100&\vr11001111\vr&00000000\vr&0000)
\cr
$V_4$: &(\vr10101010&\vr10101100\vr&11000000\vr&0000)
\cr
\omit\kern 20pt&\kern 4pt\ru&\ru&\hfil\vr&\cr
$V_5$: &(00000000&\vr00001111&11111111\vr&1111)
\cr
$V_6$: &(00010001&\vr00011110&10101010\vr&0101)
\cr
\omit\kern 20pt&&\ru&\ru&\cr
}}
\goodbreak
This structure can be thought of as three overlapping blocks
of 16: two of the blocks correspond to the boxes shown
above. The third block of 16 consists of the entries which
are in vectors $V_1$, $V_2$, $V_5$, $V_6$ and in columns
$\{$3,4,7,8,11,12,15,16,17,18,23,24,25,26,27,28$\}$.

The overlaps of the three blocks of 16 are sufficiently complicated
that it is not clear by inspection whether this structure
inherits any allowed s-c labelings. However an exhaustive search
of all $2^{28}$ possibilities shows that for this structure
of 28 {\it there are no s-c labelings satisfying} \needs .
Thus in this case the KLST prescription may break down: the
assumed modular properties \klstph\ do not agree
with \ourph .
This does
not necessarily mean that there are no consistent solutions to
string theory with this real fermion spin structure,
but that one may have to go beyond the KLST construction
to derive them.

Our final result is that the original KLST construction is
consistent for a large class of spin structures which
describe real fermions, but may fail in other cases.
Just as importantly, we have also
learned that the allowed spin structures for real fermions are
quite restricted. This is not surprising from the point of
view of rational conformal field theory, but it has important
consequences for model building.

\newsec{Conclusions}

Our work suggests a number of technical issues involving real
fermionization that need further analysis. It also suggests
some valuable model building strategies that may enable us to
eventually go beyond free fermionization. Let us begin with
two technical issues which we have not yet touched on.

\bigskip
1.\thinspace {\it Supercurrent constraints.}
Given a better understanding of the real fermion conformal field theories
it is useful to state more precisely the world-sheet supersymmetry
constraints
necessary for obtaining Lorentz invariance and $N$$=$$1$ spacetime
supersymmetry. The supercurrent of the $(1,0)$ internal superconformal
field theory of central charge $c$$=$$9$ takes the triplet form \klt \abk ,
\eqn\tripletc{
T_F ( \zbar ) ~=~ i\sum_{k=1}^{6} \, \psi_{3k} \psi_{3k+1} \psi_{3k+2}
{}~~~~~,
}
\noindent where
the $\psi_i(\zbar )$, $i$$=$$ 3, \ldots 20 $,
are right-moving Majorana-Weyl fermions,
grouped into six triplets.

Following \klt\ we will consistently choose
the internal conformal field theory part of
the spacetime supersymmetry currents to be embedded in the
tensor product of the six individual Ramond ground states
associated with $\psi_3$, $\psi_6$, $\psi_9$, $\psi_{12}$,
$\psi_{15}$, and $\psi_{18}$.
\noindent The related $U(1)$ current is the fermion bilinear
\eqn\susyc{
j(\zbar )  ~~=~~ i  \psi_3 \psi_6 +i  \psi_9 \psi_{12}
+i \psi_{15} \psi_{18}
{}~~~,}
\noindent which generates a $(2,0)$ extension of the world-sheet
superconformal algebra \bdfm . Thus, the supercurrent
\tripletc\ can be split into $T_F^+$ and $T_F^-$ as follows:
\eqn\supc{
\eqalign{
T_F^{\pm} ( \zbar ) ~~=~~ {1\over\sqrt{2}} \sum_{k=1}^{3}\,
&i\left[ \psi_{6k-3} \psi_{6k-2} \psi_{6k-1}
+ \psi_{6k} \psi_{6k+1} \psi_{6k+2} \right] \cr
&\pm\left[ \psi_{6k-2} \psi_{6k-1} \psi_{6k}
- \psi_{6k-3} \psi_{6k+1} \psi_{6k+2} \right]
\qquad .
}}
The $U(1)$ current algebra is an independent constraint
on the Hilbert space of
a consistent solution to string theory beyond
the constraints from $(1,0)$
world-sheet supersymmetry alone. Thus the superconformal constraints on
the basis vectors in a model with spacetime supersymmetry are
\eqn\triplet{
\eqalign{
&r_{6k-3} + r_{6k-2} +r_{6k-1} = r_{6k} + r_{6k+1} + r_{6k+2}
= r_{6k-2} + r_{6k-1} + r_{6k}\cr
=\,&r_{6k-3} + r_{6k+1} + r_{6k+2}
{}~=~ r_1 = r_2
{}~~~ {\rm mod ~ 1} \quad {\rm for}~k=1,2,3~~~.
}}
\noindent Here, $r_i$ denote the i'th right-moving component of any basis
vector.
This is not the usual form of the triplet
constraint stated in the literature \klt ,
but it is equivalent in any modular invariant spacetime
supersymmetric model.

If we restrict ourselves to antiperiodic and periodic boundary conditions
alone for the right-moving fermions, the superconformal conditions
\triplet\ are sufficient to guarantee a consistent
solution to string theory, assuming that the spectrum also satisfies
the modular invariance constraints. We have seen in the previous section
that this requires, in addition to \triplet , that we clearly identify
every right-moving Majorana-Weyl fermion
as either being globally paired with
a right/left-moving Majorana-Weyl fermion to
form a Weyl/Ising fermion, or as a
member of a {\it valid spin structure block}
of unpaired (right-moving and/or
left-moving) real fermions. For this class of solutions, we now have an
unambiguous prescription to build fully consistent solutions
to string theory whose underlying conformal field theory description
includes {\it both} unpaired and paired Majorana-Weyl fermions. The
two examples given in section 4 were particularly simple examples of
this class, since all of the real fermions were left-movers.
We will develop
this class of solutions in future work. In particular, it is
possible to systematically explore the options for obtaining three
generations compatible with the gauge symmetry being
realized at higher level.

It is more difficult
to implement the supercurrent constraints
for general models containing a combination of
Weyl, Ising, and real fermions.
This is because we have the possibility of
introducing twisted boundary conditions
other than periodic or antiperiodic for some of the
right-moving Weyl fermions.
In this case the supercurrent constraints require that,
{\it up to an overall basis change}
of the right-moving fermions, the boundary conditions
in the basis vectors
$\{V_i\}$ describe a set of commuting automorphisms/antiautomorphisms
of the supercurrent \abk \klt .
A detailed discussion with many examples is given in \dreiner .
An explicit prescription analogous to \triplet\ for
determining whether a {\it given} set of boundary conditions
is valid has not been derived,
and thus this class of
solutions will require further
analysis.\myfoot{In particular, we believe that
world-sheet supersymmetry
is violated for the three generation
model presented in \us .}

\vskip .2in
2.\thinspace{\it Verification.} As noted, we have developed a symbolic
manipulation package \docspec\ to analyze models constructed using
real fermionization. The program constructs the massless physical
spectrum explicitly, by solving, for every sector, the constraint
equations which implement the GSO projections.
The algorithm for solving these equations is fairly involved
due to the complicated form of the GSO projection operators for
real fermions \klst , which include products of pseudo-complexified
Ramond zero mode operators.

The results so obtained are of little use
unless we can also develop some convincing means for verification
--- both of the computer program and of the detailed algorithms which
the program implements. Fortunately there are some powerful overall physics
consistency checks at our disposal. For example, neither the program
nor the underlying algorithm ``knows'' about spacetime supersymmetry
or gauge invariance. Thus a strong physics
consistency check is to verify that all of the derived states
in the massless spectrum assemble into appropriate supermultiplets
and gauge multiplets.

However we want to stress that no amount
of checking of {\it a single model}
will ever be sufficient for verification of the results.
It is essential, in addition, to run dozens (or hundreds) of test models
with the same program, purposely attempting to generate ``peculiar''
results which signal either bugs in the code or problems
with the algorithm.
These test
models utilize spin structures that correspond to
convoluted fermionic realizations of various gauge groups and/or
extended spacetime
supersymmetry. These solutions may not be of direct physical interest but
are absolutely essential for gaining confidence in
our detailed implementation of string consistency.
Verification thus becomes the most time-consuming
aspect of building models with free fermionization.

\vskip .2in
Free fermionization is a useful paradigm for understanding how
a successful string unification model {\it might} work. There are
valuable lessons to be gained from an in-depth understanding
of this very basic tool in string theory. Of course
free fermionization has its limitations.
The restriction to constructing
solutions which realize only those gauge groups and representations
that have a fermionic embedding implies that one must be careful
in interpreting the results.
It is essential to
have the freedom to vary the underlying constituent
conformal field theories in order to avoid concluding that
a desired phenomenological outcome is ``impossible''.

On the other hand, real fermionization allows us to sample
many interesting solutions to string theory in a
calculable framework.
Realizing the world-sheet operator algebras in simpler constituents
such as free fields provides important technical advantages.
Rather than imposing modular
invariance directly on the tensor product of characters under
the necessary operator algebras, such as
current or coset algebras, we implement the much simpler task of
imposing modular invariance on the tensor product of Virasoro
characters of the constituents. Furthermore,
since the emission vertices of spacetime fields are realized
in the primary fields of the constituent conformal
field theories, their correlation functions --
which define the couplings in the superpotential -- are given by the
tensor product of constituent conformal field theory correlators.

Since our interest is {\it not} in exhaustively classifying solutions
to string theory but rather in
identifying solutions which offer new physical
insight, this repackaging of the problem will give us the capability to
efficiently access phenomenologically distinct solutions.
Already we can make a number of intriguing observations about
phenomenological properties of real fermionization.
We have identified a large number of new embeddings
of GUT groups and the standard model group, realized at higher level.
The two examples presented here demonstrate that different choices
of embeddings lead to quite different particle content in the effective
field theory. We find that a limited number of adjoint scalars, other
large Higgs irreps, and exotics can appear in our models, with highly
model dependent couplings. The number of gauge-singlet moduli can also
be quite small, a result which may have important phenomenological
consequences. There are interesting new possibilities for the hidden
sector gauge group and matter content. Last but not least, real
fermionization clearly restricts the operators that give fermions mass
in ways that differ strongly from previous constructions.

It might seem that,
given a sufficiently wide range of constituent conformal field theories,
anything and everything is possible in the spectrum and in the
superpotential.
This is a
misconception. As we have repeatedly emphasized, and as is evident in
any experience with building explicit solutions, string consistency
is a very restrictive principle. Slight changes in the underlying conformal
field theory embeddings can have rather drastic consequences for the
massless spectrum and the superpotential. Given the dictionary
between spacetime symmetries and world sheet operator algebras, it is
probably not difficult to construct
conformal field theory structures
that realize any {\it single} phenomenological feature,
assuming it satisfies
the bounds on allowed conformal dimension and total conformal
anomaly \dkl \fer . But the final step of piecing together
{\it many} features in a consistent solution is extremely delicate.
It is this property which makes superstring unification so restrictive,
but also compelling.

\centerline{\bf ACKNOWLEDGEMENTS}

We would like to thank G. Cleaver, A. Faraggi,
J. Feng, D. Finnell, Z. Kakushadze,
J. Lopez, M. Peskin, S. Raby, P. Ramond,
H. Tye, and K. Yuan for discussions and useful information.
S.C. also thanks T. Banks, K. Dienes,
L. Dixon, A. Nelson, L. Randall, S. Shenker, and A. Strominger.
S.C. is indebted to Joe
Polchinski for many stimulating discussions and insights.
This work was supported by the U.S. Department of Energy
under contract DE-AC02-76CHO3000.
The work of S.C. is
supported by National Science
Foundation grants PHY-91-16964 and PHY-89-04035.

\listrefs
\goodbreak
\vglue .3in
\leftline{\bf TABLE CAPTIONS}
\vskip .3in
\noindent{\bf Table 1:} The eight sectors which contribute the
46 gauge bosons of $SO(10)$$\times$$U(1)$ in Model A.
\vskip .3in
\noindent{\bf Table 2:} The fermionic charges of the 46 gauge bosons
of $SO(10)$$\times$$U(1)$ in Model A, listed according
to the sectors that they appear in.
\vskip .3in
\noindent{\bf Table 3:} The complete massless spectrum of chiral
superfields for Model A. A $\pm$ indicates two distinct irreps with
opposite charge: thus, for example, there are a total of four 16's of
$SO(10)$ and a total of twelve 10's of $SO(10)$.
\vskip .3in
\noindent{\bf Table 4:} The eight sectors which contribute the
45 scalars of the adjoint Higgs in Model A.
\vskip .3in
\noindent{\bf Table 5:} The eleven sectors which contribute the
45 gauge bosons of $SO(10)$ in Model B.
\vskip .3in
\noindent{\bf Table 6:} The fermionic charges of the 45 gauge bosons
of $SO(10)$ in Model B, listed according to the sectors that they
appear in.
\vskip .3in
\noindent{\bf Table 7:} The fermionic charges of the scalars of
the 54 of $SO(10)$ in Model B. The ellipsis indicates that the
remaining entries are identical to the those in Table 6.
\vskip 2in
\goodbreak
\vglue .3in
\halign to \hsbody{\kern 10pt$#$\hfil&\kern 2em#&\kern 1em#\cr
\omit\kern 10pt
\underbar{Sector}&\underbar{No. of gauge boson states}
&\underbar{Real fermion b.c.'s}\cr
\omit\kern 10pt untwisted&10&(0000000000000000)\cr
V_2&8&(1111111100000000)\cr
V_3&4&(0000111111110000)\cr
V_4&4&(1100110011001100)\cr
V_2$$+$$V_3&8&(1111000011110000)\cr
V_2$$+$$V_4&4&(0011001111001100)\cr
V_3$$+$$V_4&4&(1100001100111100)\cr
V_2$$+$$V_3$$+$$V_4&4&(0011110000111100)\cr
}
\vskip .3in\centerline{Table 1}
\vfil\eject
\vglue .3in
\halign to \hsbody{$#$\hfil&\kern 3em#&\kern 1em#\cr
\omit
\underbar{Sector}&
\underbar{Fermionic charges}&\cr
\omit untwisted:&5$\times$(0,0,0,0,0,0)&\cr
&$\pm$(1,0,0,0,0,0)&$\pm$(0,1,0,0,0,0)\cr
V_2:&$\pm$(1/2,-1/2,1/2,-1/2,0,0)&$\pm$(1/2,-1/2,-1/2,1/2,0,0)\cr
&$\pm$(1/2,1/2,1/2,-1/2,0,0)&$\pm$(1/2,1/2,-1/2,1/2,0,0)\cr
V_3:&$\pm$(0,0,1,0,0,0)&$\pm$(0,0,0,1,0,0)\cr
V_4:&$\pm$(1/2,0,0,1/2,1/2,-1/2)&$\pm$(1/2,0,0,-1/2,-1/2,1/2)\cr
V_2$$+$$V_3:&$\pm$(1/2,-1/2,-1/2,-1/2,0,0)&$\pm$(1/2,-1/2,1/2,1/2,0,0)\cr
&$\pm$(1/2,1/2,1/2,1/2,0,0)&$\pm$(1/2,1/2,-1/2,-1/2,0,0)\cr
V_2$$+$$V_4:&$\pm$(0,1/2,1/2,0,1/2,-1/2)&$\pm$(0,1/2,-1/2,0,-1/2,1/2)\cr
V_3$$+$$V_4:&$\pm$(1/2,0,0,-1/2,1/2,-1/2)&$\pm$(1/2,0,0,1/2,-1/2,1/2)\cr
V_2$$+$$V_3$$+$$V_4:&$\pm$(0,1/2,-1/2,0,1/2,-1/2)&$\pm$(0,1/2,1/2,0,-1/2,1/2)
\cr
}
\vskip .3in\centerline{Table 2}
\vfil\eject
\vglue .2in
\halign to \hsbody{\kern 10pt\hfil$#$\hfil&
\kern 2em\hfil#\hfil&
\kern 1em\hfil#&\hfil#&\hfil#&\hfil#\cr
\omit\kern 10pt
\underbar{Irrep of $SO(10)$$\times$$SO(8)$}&
\underbar{Multiplicity}&&
\hfil\underbar{$U(1)$ charges}\span\span\cr
45&2&0&0&0&0\cr
16&1&$\pm$1/2&0&-1/4&0\cr
16&2&0&0&-1/4&0\cr
\overline{16}&2&0&0&1/4&0\cr
10&2&$\pm$1/2&1/2&0&0\cr
10&1&$\pm$1/2&-1/2&0&0\cr
10&2&0&$\pm$1/2&0&0\cr
10&1&0&0&$\pm$1/2&0\cr
8_v&1&0&-1/2&-1/2&0\cr
8_s&1&0&1/2&-1/2&0\cr
8_c&1&0&1/2&1/2&0\cr
1&3&$\pm$1&0&0&0\cr
1&1&$\pm$1&-1/2&1/2&0\cr
1&1&$\pm$1/2&1/2&1/2&0\cr
1&1&$\pm$1/2&1/2&-1/2&0\cr
1&2&$\pm$1/2&-1/2&1/2&0\cr
1&2&$\pm$1/2&-1/2&-1/2&0\cr
1&1&0&$\pm$1&0&0\cr
1&2&0&$\pm$1/2&1/2&0\cr
1&1&0&1/2&-1/2&$\pm$1/2\cr
1&2&0&$\pm$1/2&-1/2&0\cr
1&1&0&0&0&$\pm$1/2\cr
1&7&0&0&0&0\cr
}
\vskip .3in\centerline{Table 3}
\vfil\eject
\vglue .3in
\halign to \hsbody{\kern 10pt$#$\hfil&\kern 2em#&\kern 1em#\cr
\omit\kern 10pt
\underbar{Sector}&\underbar{No. of states}
&\underbar{Real fermion b.c.'s}\cr
V_8&9&(0101010101010101)\cr
V_2$$+$$V_8&8&(1010101001010101)\cr
V_3$$+$$V_8&4&(0101101010100101)\cr
V_4$$+$$V_8&4&(1001100110011001)\cr
V_2$$+$$V_3$$+$$V_8&8&(1010010110100101)\cr
V_2$$+$$V_4$$+$$V_8&4&(0110011010011001)\cr
V_3$$+$$V_4$$+$$V_8&4&(1001011001101001)\cr
V_2$$+$$V_3$$+$$V_4$$+$$V_8&4&(0110100101101001)\cr
}
\vskip .3in\centerline{Table 4}
\vfil\eject
\vglue .3in
\ifx\answ\bigans\vskip .1in\else\vskip .2in\fi
\goodbreak
\halign to \hsbody{\kern 10pt$#$\hfil&\kern 2em#&\kern 1em#\cr
\omit\kern 10pt
\underbar{Sector}&\underbar{No. of gauge boson states}
&\underbar{Real fermion b.c.'s}\cr
\omit\kern 10pt untwisted&5&(0000000000000000)\cr
V_2&4&(1111111100000000)\cr
V_3&4&(1111000011110000)\cr
V_4&4&(1111000000001111)\cr
V_5&4&(0000000000000000)\cr
V_2$$+$$V_3&4&(0000111111110000)\cr
V_2$$+$$V_4&4&(0000111100001111)\cr
V_2$$+$$V_5&4&(1111111100000000)\cr
V_3$$+$$V_4&4&(0000000011111111)\cr
V_3$$+$$V_5&4&(1111000011110000)\cr
V_4$$+$$V_5&4&(1111000000001111)\cr
}
\vskip .3in\centerline{Table 5}
\vfil\eject
\vglue .3in
\halign to \hsbody{$#$\hfil&\kern 1em#&\kern 1em#\cr
\omit
\underbar{Sector}&
\underbar{Fermionic charges}&\cr
\omit untwisted:&5$\times$(0,0,0,0,0,0,0,0,0,0)&\cr
V_2:&$\pm$(1/2,1/2,1/2,1/2,0,0,0,0,0,0)&$\pm$(1/2,1/2,-1/2,-1/2,0,0,0,0,0,0)
\cr
V_3:&$\pm$(1/2,1/2,0,0,1/2,1/2,0,0,0,0)&$\pm$(1/2,1/2,0,0,-1/2,-1/2,0,0,0,0)
\cr
V_4:&$\pm$(1/2,1/2,0,0,0,0,1/2,1/2,0,0)&$\pm$(1/2,1/2,0,0,0,0,-1/2,-1/2,0,0)
\cr
V_5:&$\pm$(1/2,1/2,0,0,0,0,0,0,1/2,1/2)&$\pm$(1/2,1/2,0,0,0,0,0,0,-1/2,-1/2)
\cr
V_2$$+$$V_3:&$\pm$(0,0,1/2,1/2,1/2,1/2,0,0,0,0)&$\pm
$(0,0,1/2,1/2,-1/2,-1/2,0,0,0,0)
\cr
V_2$$+$$V_4:&$\pm$(0,0,1/2,1/2,0,0,1/2,1/2,0,0)&$\pm
$(0,0,1/2,1/2,0,0,-1/2,-1/2,0,0)
\cr
V_2$$+$$V_5:&$\pm$(0,0,1/2,1/2,0,0,0,0,1/2,1/2)&$\pm
$(0,0,1/2,1/2,0,0,0,0,-1/2,-1/2)
\cr
V_3$$+$$V_4:&$\pm$(0,0,0,0,1/2,1/2,1/2,1/2,0,0)&$\pm
$(0,0,0,0,1/2,1/2,-1/2,-1/2,0,0)
\cr
V_3$$+$$V_5:&$\pm$(0,0,0,0,1/2,1/2,0,0,1/2,1/2)&$\pm
$(0,0,0,0,1/2,1/2,0,0,-1/2,-1/2)
\cr
V_4$$+$$V_5:&$\pm$(0,0,0,0,0,0,1/2,1/2,1/2,1/2)&$\pm
$(0,0,0,0,0,0,1/2,1/2,-1/2,-1/2)
\cr
}
\vskip .3in\centerline{Table 6}
\vfil\eject
\vglue .5in
\halign to \hsbody{$#$\hfil&\kern 1em#&\kern 1em#\cr
\omit
\underbar{Sector}&
\underbar{Fermionic charges}&\cr
\omit untwisted:&4$\times$(0,0,0,0,0,0,0,0,0,0)&\cr
&$\pm$(1,1,0,0,0,0,0,0,0,0)&$\pm$(0,0,1,1,0,0,0,0,0,0)\cr
&$\pm$(0,0,0,0,1,1,0,0,0,0)&$\pm$(0,0,0,0,0,0,1,1,0,0)\cr
&$\pm$(0,0,0,0,0,0,0,0,1,1)&\cr
\ldots&&\cr
}
\vskip .3in\centerline{Table 7}
\vskip .2in

\end